\documentclass{aa}  

\usepackage{graphicx}
\usepackage{txfonts}
\usepackage{natbib}
\bibpunct{(}{)}{;}{a}{}{,} 
\usepackage{color}
\usepackage{float}
\usepackage{subfigure}
\usepackage{hyperref}
\usepackage{longtable}
\usepackage{rotating}
\usepackage{lscape}
\usepackage{soul}
\usepackage{placeins}

\newcommand{\udft}{\textsf{udf-10}}
\newcommand{\mosaic}{\textsf{mosaic}}
\newcommand{\mxdf}{\textsf{mxdf}}

\newcommand{\flcgs}{\mbox{erg s$^{-1}$ cm$^{-2}$}}
\newcommand{\sbl}{\mbox{erg s$^{-1}$ cm$^{-2}$ arcsec$^{-2}$}}

\def\HI{\mbox{H\,{\sc i}}}
\newcommand{\OII}{[O\,{\sc ii}]}

\newcommand{\MgII}{Mg\,{\sc ii}}
\newcommand{\FeII}{Fe\,{\sc ii}*}
\newcommand{\SiIIf}{Si\,{\sc ii}*}
\newcommand{\SiII}{Si\,{\sc ii}}
\newcommand{\CIIf}{C\,{\sc ii}*}
\newcommand{\CII}{[C\,{\sc ii}]}

\newcommand{\lya}{\mbox{Ly$\alpha$}}
\begin{document} 

   \title{The MUSE eXtremely Deep Field: Detections of circumgalactic \SiIIf\ emission at $z\gtrsim2$\thanks{Based on observations made with ESO telescope at the La Silla
Paranal Observatory under the large program 1101.A-0127}}


 \author{Haruka Kusakabe\inst{\ref{inst2},\ref{inst1}\thanks{e-mail: haruka.kusakabe.takeishi@gmail.com, haruka.kusakabe@nao.ac.jp}}
 \and Valentin Mauerhofer\inst{\ref{inst3}}
 \and Anne Verhamme\inst{\ref{inst2}}
 \and Thibault Garel \inst{\ref{inst2}}
 \and J{\'e}r{\'e}my Blaizot\inst{\ref{inst4}}  
 \and Lutz Wisotzki\inst{\ref{inst5}}
 \and Johan Richard\inst{\ref{inst4}}
 \and Leindert A. Boogaard\inst{\ref{inst6}}
 \and Floriane Leclercq\inst{\ref{inst7}}
 \and Yucheng Guo\inst{\ref{inst4}}
 \and Ad{\'e}la\"{i}de Claeyssens\inst{\ref{inst8}}
 \and Thierry Contini\inst{\ref{inst9}}
 \and Edmund Christian Herenz\inst{\ref{inst10}}
 \and Josephine Kerutt\inst{\ref{inst3}}
 \and Michael V. Maseda\inst{\ref{inst11}}
 \and Leo Michel-Dansac\inst{\ref{inst4},\ref{inst12}}
\and Themiya Nanayakkara\inst{\ref{inst13}}
\and Masami Ouchi\inst{\ref{inst1},\ref{inst14},\ref{inst15},\ref{inst16}} 
 \and Ismael Pessa\inst{\ref{inst5}}
 \and Joop Schaye\inst{\ref{inst17}} 
}
\institute{
Observatoire de Gen\`{e}ve, Universit\'e de Gen\`{e}ve, 51 Chemin de P\'egase, 1290 Versoix, Switzerland\label{inst2}
\and%
National Astronomical Observatory of Japan, 2-21-1 Osawa, Mitaka, Tokyo, 181-8588, Japan\label{inst1}
\and
Kapteyn Astronomical Institute, University of Groningen, P.O Box 800, 9700 AV Groningen, The Netherlands\label{inst3}
\and
Univ. Lyon, Univ. Lyon1, Ens de Lyon, CNRS, Centre de Recherche Astrophysique de Lyon UMR5574, F-69230, Saint-Genis-Laval, France\label{inst4}
\and
Leibniz-Institut f\"{u}r Astrophysik Potsdam (AIP), An der Sternwarte 16 14482 Potsdam, Germany\label{inst5}
\and Max Planck Institute for Astronomy, K\"{o}nigstuhl 17, 69117 Heidelberg, Germany\label{inst6}
\and 
Department of Astronomy, The University of Texas at Austin, 2515 Speedway, Stop C1400, Austin, TX 78712-1205, USA\label{inst7}
\and 
The Oskar Klein Centre, Department of Astronomy, Stockholm University, AlbaNova, SE-10691 Stockholm, Sweden\label{inst8}
\and 
Institut de Recherche en Astrophysique et Plan, Toulouse 14, avenue E. Belin, 31400, France\label{inst9}
\and 
Inter-University Centre for Astronomy and Astrophysics, Ganeshkind, Post Bag 4, Pune 41007, India\label{inst10}
\and 
Department of Astronomy, University of Wisconsin-Madison, 475 N. Charter St., Madison, WI 53706, USA\label{inst11}
\and
Aix Marseille Univ, CNRS, CNES, LAM, Marseille, France\label{inst12}
\and
Centre for Astrophysics and Supercomputing, Swinburne University of Technology, Hawthorn, Victoria 3122, Australia\label{inst13}
\and
Institute for Cosmic Ray Research, The University of Tokyo, 5-1-5 Kashiwanoha, Kashiwa, Chiba 277-8582, Japan\label{inst14}
\and
Kavli Institute for the Physics and Mathematics of the Universe (WPI), University of Tokyo, Kashiwa, Chiba 277-8583, Japan\label{inst15}
\and
Department of Astronomical Science, SOKENDAI (The Graduate University for Advanced Studies), Osawa 2-21-1, Mitaka, Tokyo, 181-8588, Japan\label{inst16}
\and Leiden Observatory, Leiden University, PO Box 9513, NL-2300 RA Leiden, The Netherlands\label{inst17}
}
  \abstract
   {The circumgalactic medium (CGM) serves as a baryon reservoir that connects galaxies to the intergalactic medium and fuels star formation. The spatial distribution of the metal-enriched cool CGM has not yet been directly revealed at cosmic noon ($z\simeq2$--$4$), as bright emission lines at these redshifts are not covered by optical integral field units. }
   {To remedy this situation, we aim for the first-ever detections and exploration of extended \SiIIf\ emission (low-ionization state, LIS), referred to as ``\SiIIf\ halos'', at redshifts ranging from $z=2$ to $4$ as a means to trace the metal-enriched cool CGM.}
   {We use a sample of 39 galaxies with systemic redshifts of $z=2.1$--$3.9$ measured with the [C\,{\sc iii}] doublet in the MUSE Hubble Ultra Deep Field catalog, whose integration times span from $\simeq30$ to $140$ hours. We search for extended \SiIIf\ $\lambda$1265, 1309, 1533 emission (fluorescent lines) around individual galaxies. We also stack a subsample of 14 UV-bright galaxies.}
   {We report five individual detections of \SiIIf\ $\lambda1533$ halos. We also confirm the presence of \SiIIf\ $\lambda1533$ halos in stacks for the subsample containing UV-bright sources. The other lines do not show secure detections of extended emission in either individual or stacking analyses. These detections may imply that the presence of metal-enriched CGM is a common characteristic for UV-bright galaxies. 
   To investigate whether the origin of \SiIIf\ is continuum pumping as suggested in previous studies, we check the consistency of the equivalent width (EW) of \SiIIf\ emission and the EW of \SiII\ absorption for the individual halo object with the most reliable detection. We confirm the equivalence, suggesting that photon conservation works for this object and pointing toward continuum pumping as the source of \SiIIf. We also investigate \SiIIf\ lines in a RAMSES-RT zoom-in simulation including continuum pumping and find ubiquitous presence of extended halos. 
   }
   {}
   \keywords{Galaxies: high-redshift -- galaxies: formation -- galaxies: evolution -- galaxies: halos 
               }
   \maketitle
%

\section{Introduction}\label{sec:intro}
\begin{figure*}
   \includegraphics[width=1.0\hsize]{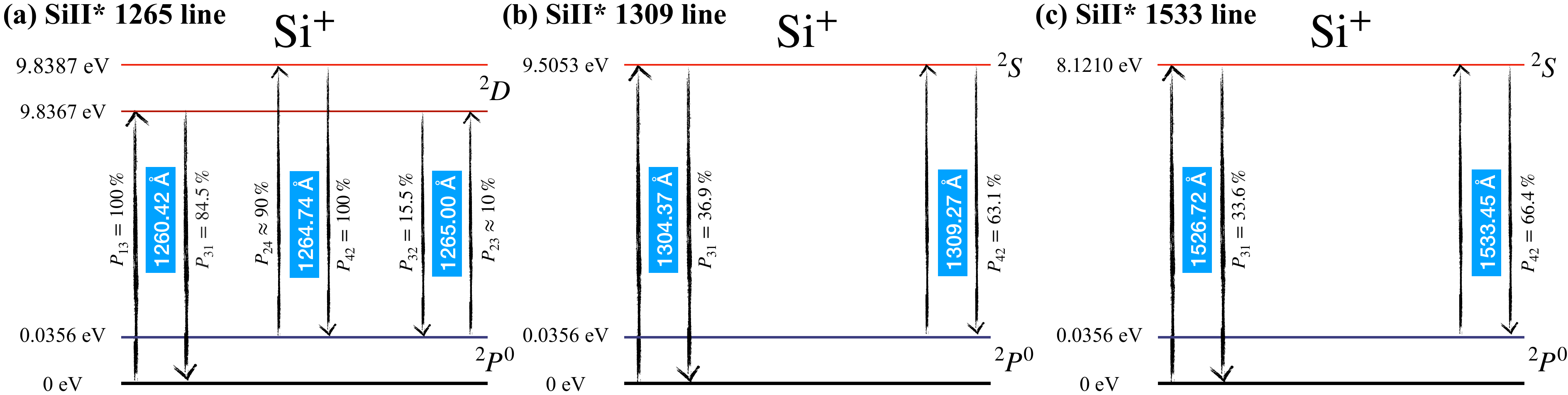}
      \caption{ Energy levels of Si$^+$ ions and different channels of excitation and de-excitation from the ground-state or the fine-structure level (fluorescence) for \SiIIf\ $\lambda$1265, 1309, 1533 lines in panels (a) to (c), respectively. The probabilities ($P$) of related transitions and their wavelengths are also shown. The ionization energies of these LIS lines are close to that of \HI. The \SiIIf\ lines can be regarded as an escape channel of photons from \SiII\ resonant scattering. 
      }
         \label{fig:energy_levels}
\end{figure*}

The circumgalactic medium (CGM) is the baryonic matter that envelops galaxies. The most common definition of the CGM is the gas outside the interstellar medium (ISM) and inside the virial radius of the host dark matter halo. This region serves as a dynamic interface connecting galaxies to the intergalactic medium (IGM). Metals are ejected from galaxies to the CGM via outflows powered by supernovae explosions, stellar winds, and active galactic nuclei (AGNs). A part of those metals may be mixed with inflowing pristine and recycled gas and accrete onto the galaxies. Hence, the CGM is a gas and metal reservoir of a galaxy and plays a pivotal role in galaxy evolution \citep[e.g.,][]{tumlinson_circumgalactic_2017, peroux_cosmic_2020}.

Historically, the CGM has been observed in absorption imprinted on quasi-stellar object (QSO) spectra \citep[e.g.,][]{wolfe_damped_1986}. This method can trace a wide range of column densities and ionization states of various elements. The CGM is found to be a multiphase medium in terms of its density, temperature, ionization, kinematics, metallicity, and structure \citep[e.g.,][]{werk_cos-halos_2016,steidel_reconciling_2016, chen_cosmic_2020,schroetter_muse_2021}. However, this method is limited to the lines of sight of background sources. Large programs with recent or upcoming instruments targeting bright galaxies as background sources (tomographic mapping) can reach the IGM scale. For instance, The COSMOS Lyman-Alpha Mapping And Tomography Observations (CLAMATO) survey with Keck/LRIS \citep[][]{lee_ly_2014} and the Prime Focus Spectrograph-Subaru strategic program (PFS SSP) Galaxy Evolution survey \citep[][]{greene_prime_2022} have a few Mpc transverse resolution at $z>1$. Higher spatial resolutions can be achieved using QSO pairs \citep[e.g., ][]{tytler_metal_2009,urbano_stawinski_metallicities_2023}, gravitationally lensed QSOs \citep[e.g., ][]{ chen_spatially_2014,rubin_andromedas_2018}, and gravitational-arcs as background sources \citep[e.g., ][]{lopez_clumpy_2018}. A galaxy-centered stacking approach is also useful to statistically map gas and metals with impact parameters ranging from $\simeq50$ kpc to a few Mpc \citep[e.g., ][]{rakic_neutral_2012,turner_metal-line_2014, dutta_musequbes_2024}. Recently, \citet{bordoloi_eiger_2023} achieves the impact parameters of $\simeq20$--$300$ kpc individually, using {\it James Web} Space Telescope (JWST)/NIRCam slitlesss grism spectroscopy. These studies enable to reach an outer-CGM scale, but an inner-CGM scale\footnote{Roughly within radii of $R\lesssim0.2 R_{\rm vir}$, where $R_{\rm vir}$ is virial radius. The inner CGM is defined as $R<50$ kpc in \citet{tumlinson_circumgalactic_2017} for L$^*$ galaxies at $z=0.2$ \citep[with a median $R_{\rm vir}\simeq 320$ kpc in][]{werk_cos-halos_2014}.} is difficult to investigate. Moreover, the identification of the host galaxies that are responsible for the absorption lines in QSO spectra is a challenging task. Hence, this method cannot provide the spatial distribution of gas and metals around individual host galaxies on an inner CGM scale.

A complementary method to absorption-line studies is direct observations of the CGM in emission, which has been challenging because of its low surface brightness (SBs). Modern optical integral field units (IFUs) such as the Multi-Unit Spectroscopic Explorer \citep[MUSE;][]{bacon_muse_2010} and the Keck Cosmic Web Imager \citep[KCWI;][]{morrissey_keck_2018} make it possible to detect diffuse CGM emission around individual host galaxies \citep[e.g., ][]{wisotzki_extended_2016}. In particular, extended Ly$\alpha$ emission, the so-called ``Ly$\alpha$ halo'', which traces cool and warm hydrogen CGM \citep[e.g.,][]{steidel_diffuse_2011,momose_diffuse_2014,guo_median_2024} has been intensively detected and studied around individual Ly$\alpha$ emitters (LAEs) at $z\simeq2$--$6$ \citep[e.g.,][]{leclercq_muse_2017,leclercq_muse_2020, chen_kbsskcwi_2021,claeyssens_lensed_2022,erb_circumgalactic_2023}. Their diverse profiles and statistical trends as well as origins can now be discussed. Recently, a high fraction of Ly$\alpha$ halos around UV-continuum selected (not based on Ly$\alpha$) galaxies at $z\sim3$-$4$ is reported \citep[][]{kusakabe_muse_2022}. The ubiquitous presence of resorvoirs of hydrogen in the CGM is directly confirmed at cosmic noon and earlier epochs ($z>3$ with MUSE).  

However, metals in the CGM at cosmic noon ($z\sim2$--$4$), which is the peak of star-forming activity of galaxies, have not been directly confirmed in emission, in particular for the cool phase. The CGM is known to be metal enriched even at $z\sim2$--$4$ from transverse absorption-line studies \citep[e.g.,][]{turner_metal-line_2014, lehner_cosmic_2016, lehner_kodiaq-z_2022, mendez-hernandez_metal_2022,urbano_stawinski_metallicities_2023,bordoloi_eiger_2024, banerjee_musequbes_2023, beckett_muse_2024}. This is also suggested by outflowing metals identified by down-the-barrel analysis \citep[e.g.,][]{shapley_rest-frame_2003, steidel_structure_2010,du_redshift_2018,sugahara_evolution_2017} and by broad-line components seen in spectra \citep[e.g.,][]{carniani_jades_2024,xu_stellar_2023}. A popular emission tracer of cool metal-enriched CGM, \MgII\ $\lambda\lambda$2796, 2803, is useful only at $z\lesssim2$ because at higher redshifts the observed wavelength shifts into the near-infrared and the cosmic dimming effect is more severe \citep[e.g.,][]{rubin_low-ionization_2011, erb_galactic_2012, martin_scattered_2013, burchett_circumgalactic_2021, zabl_muse_2021, leclercq_muse_2022, dutta_metal_2023, guo_bipolar_2023,pessa_galactic_2024}. Other possible tracers are \OII$\lambda\lambda3726, 3729$ \citep[e.g.,][]{yuma_first_2013, yuma_systematic_2017, epinat_ionised_2018, johnson_galaxy_2018} and very faint \FeII$\lambda2365$, $\lambda2396$, $\lambda2612$, and $\lambda2626$ \citep[e.g.,][]{finley_galactic_2017,shaban_30_2022}. At $z\simeq4$--$7$, \CII$\lambda$158$\mu$m can be covered by sensitive ALMA bands 6 and 7 \citep[\CII\ halos; e.g.,][]{fujimoto_first_2019, ginolfi_alpine-alma_2020}. At $z\simeq2$--$4$, no bright emission tracer of the cool metal-enriched CGM is available in the observed optical regime. While extremely faint, the most promising tracers are the emission lines \SiIIf\ $\lambda$1265, 1309, 1533, which are not contaminated by other line features at nearby wavelengths given the spectral resolution of current facilities (cf. C\,{\sc ii}*$\lambda$1335 emission and C\,{\sc ii}$\lambda$1334 absorption, which cannot be resolved by the resolving power of MUSE, $R=1770$--$3590$ for $\lambda=4650$--$9350$\AA, or $\Delta\lambda\simeq3$~\AA\ for \CIIf$\lambda1335$ line at $z=2.5$).  

\SiIIf\ are fine-structure fluorescent emission lines from singly ionized silicon, which is a low-ionization state (LIS), with an ionization energy close to \HI\ (ionization energy for Si and Si$^{+}$ are 8.1~ eV and 16.4~eV, respectively). The energy levels of the Si$^{+}$ ion for the three transitions this paper focuses on are shown in Fig. \ref{fig:energy_levels}. The origin of \SiIIf\ is suggested to be ``continuum pumping'' (or ``continuum fluorescence'') rather than collisional excitation or recombination \citep[][]{shapley_rest-frame_2003}. Collisional excitations are dominant only in dense environments with high electron densities, which is higher than typical values in H\,{\sc ii} regions \citep{shapley_rest-frame_2003}. And the recombination rates of Si$^{2+}$ into the excited Si$^{+}$ is of the same order as the collisional excitation rates when Si$^{+}$ and Si$^{2+}$ have comparable abundance with $T\sim10^4$ K \citep{shull_ionization_1982, shapley_rest-frame_2003}. \citet{shapley_rest-frame_2003} model observed nebular emission lines using the photoionization code CLOUDY \citep{ferland_cloudy_1998}. They find that any model that reproduces the line ratios of the other lines ([O\,{\sc iii}], [O\,{\sc ii}], H$\beta$, O\,{\sc iii}], and C\,{\sc iii}]) predicts more than one order of magnitude weaker \SiIIf\ emission by collisions and recombination. Therefore, those two processes are unlikely to be the dominant origin of \SiIIf\ emission in H\,{\sc ii} regions. As for the continuum pumping scenario, first, Si$^+$ in neutral clouds absorbs the UV continuum and creates \SiII\ resonant absorption features. Second, a fraction of Si$^+$ deexcites to the fine-structure level and emits \SiIIf\ photons (see Fig. \ref{fig:energy_levels}). \SiIIf\ lines are escape channels of resonant transitions and are expected to be as spatially extended as the resonant emission lines.

Fluorescent lines are extremely faint, and \SiIIf\ emission is $\sim10$--$50$ times weaker than Ly$\alpha$ emission of LAEs \citep[][]{steidel_keck_2018}. In fact, extended \SiIIf\ emission has not been directly detected. \citet{wang_systematic_2020} found weak \SiIIf\ emission lines compared to the associated absorption line in HST/Cosmic Origins Spectrograph (COS) data for 5 galaxies at $z=0$, which implies that the bulk of \SiIIf\ emission arises on larger scales than the COS aperture. \citet{gazagnes_interpreting_2023} compare COS spectra of \SiIIf\ $\lambda$1265 for 45 local galaxies (COS Legacy Archive Spectroscopic SurveY, CLASSY) with mock spectra from zoom-in simulations \citep{mauerhofer_uv_2021}. They find stronger \SiIIf\ emission lines in the simulations than observed, and they argue that aperture losses of COS can explain the weakness of observed \SiIIf\ and that \SiIIf\ $\lambda$1265 is spatially extended. Very recently, \citet{keerthi_vasan_g_spatially_2024} study spatially-resolved outflow properties for a lensed star-forming galaxy at $z=1.87$ with KCWI and show that \SiIIf\ emission is more extended than the continuum for a certain direction extracted with a pseudo slit. 

In this project, we search for \SiIIf\ halos around galaxies at cosmic noon (at $z=2$--$4$) as a tracer of metal-enriched cool CGM using integral field unit data to remedy the problem. We use data from the MUSE Hubble Ultra Deep Field (HUDF) survey \citep{bacon_muse_2023} and test the presence of \SiIIf\ halos with surface brightness profiles of \SiIIf\ and UV continuum. We report the first detections of \SiIIf\ halos. We also stack the MUSE data for a UV-bright subsample to study the general extent of \SiIIf\ and compare it with simulations.

The paper is organized as follows. In Section \ref{sec:data_sample}, we describe the data and the sample construction. Section \ref{sec:result} presents methods and results of the search for individual \SiIIf\ halos and those of a stacked subsample. In Section \ref{sec:discussion}, we discuss photon conservation for \SiII, compare observed results with those of zoom-in simulations, and discuss how to increase a sample size with different selection criteria. Finally, conclusions are given in Section \ref{sec:conclusions}. Throughout this paper, we assume the Planck 2018 cosmological model \citep{planck_collaboration_planck_2020} with a matter density of $\Omega_{\rm m} = 0.315$, a dark energy density of $\Omega_{\Lambda} = 0.685$, and a Hubble constant of $H_0 = 67.4$ km s$^{-1}$ Mpc$^{-1}$ ($h_{100} = 0.67$). Magnitudes are given in the AB system \citep{oke_secondary_1983}. All distances are in physical units (kpc), unless otherwise stated.

\section{Data and sample}\label{sec:data_sample}

To search for extended \SiIIf\ emission of $z\gtrsim2$ galaxies, we utilize deep MUSE data from data release 2 (DR2) of the MUSE HUDF surveys \citep{bacon_muse_2023}, where rich multi-wavelength data are also available (see Section \ref{subsec:data}). We construct our sample with the DR2 catalog using the systemic redshifts for the individual search and stacking analyses (see Sections \ref{subsec:sample} and \ref{subsec:sample_stack}). We create masks to avoid potential contaminations from neighboring galaxies (see Section \ref{subsec:sample_mask}).

\subsection{Data and catalogs}\label{subsec:data}
The MUSE HUDF DR2 data are obtained as a part of the MUSE guaranteed time observations (GTO) program described in \citet{bacon_muse_2023}. We use two fields among the MUSE HUDF DR2 data set: a single $1\times1$ arcmin$^2$ pointing with a 31-hour depth (\udft, see also \citealt{bacon_muse_2017} and \citealt{inami_muse_2017}), and the adaptive-optics (AO) assisted MUSE eXtremely Deep Field (\mxdf), whose maximum integration time is 141 hours with a 1-arcmin diameter field of view. They are located inside the {\it Hubble} eXtreme Deep Field \citep[XDF;][]{illingworth_hst_2013} with deep HST data \citep[see][for more details]{bacon_muse_2023}. The \mxdf\ is the deepest IFU survey ever performed, while the \udft\ is the second deepest available for this study, which also has multi-wavelength data and catalogs constructed in the same manner as \mxdf. The \udft\ covers the optical wavelength range from 4750~\AA\ to 9350~\AA, while \mxdf\ covers from  4700~\AA\ to 9350~\AA, with an AO gap from 5800~\AA\ to 5966.25~\AA. The spectral resolving power of MUSE varies from $R=1610$ to 3750 at 4700~\AA\ to 9350~\AA, respectively. The full width at half maximums (FWHMs) of the Moffat point spread function \citep[Moffat PSF,][]{moffat_theoretical_1969} are $0\farcs7$ ($0\farcs7$) at the blue wavelength edge and $0\farcs5$ ($0\farcs6$) at the red wavelength edge for the \mxdf\ (\udft). The $3\sigma$ point-source flux limit for an unresolved emission line in the \mxdf\ (the \udft) is $\simeq6\times10^{-20}$ \flcgs\ ($\simeq2\times10^{-19}$ \flcgs), at around 7000~\AA\ (not affected by OH sky emission), which corresponds to a $3\sigma$ surface brightness limit for an unresolved emission line of $\simeq1\times10^{-19}$ \sbl\ ($\simeq2\times10^{-19}$ \sbl, see \citealt{bacon_muse_2023}). 

The MUSE HUDF DR2 catalog includes 2221 sources from $z=0$ to $6.7$, which are selected by emission-line detections as well as through HST priors. The details of the catalog construction are described in \citet{bacon_muse_2023}. Each MUSE object has a source file, which is in the format of MPDAF\footnote{\url{https://mpdaf.readthedocs.io/en/latest/}} multi-fits format and is composed of various data files such as minicubes, images, and spectra. It is available on the AMUSED website\footnote{\url{https://amused.univ-lyon1.fr/}}. We use 5''$\times$5'' minicubes from each of the source files. We create continuum-subtracted minicubes following the method used in \citet{kusakabe_muse_2022}, taking the median for each pixel within a spectral window of 100 slices (masking $\pm400$ km s$^{-1}$ around the $\lya$ wavelength if the spectrum covers). The continuum subtraction is useful not only to investigate emission lines, but also to remove neighboring sources around a target source. In order to identify HST counterparts, calculate UV magnitudes, and create masks, we use the HST catalog and data from \citet{rafelski_uvudf_2015}.

\subsection{Sample selection}\label{subsec:sample}

\begin{figure*}
\sidecaption
   \includegraphics[width=12cm]{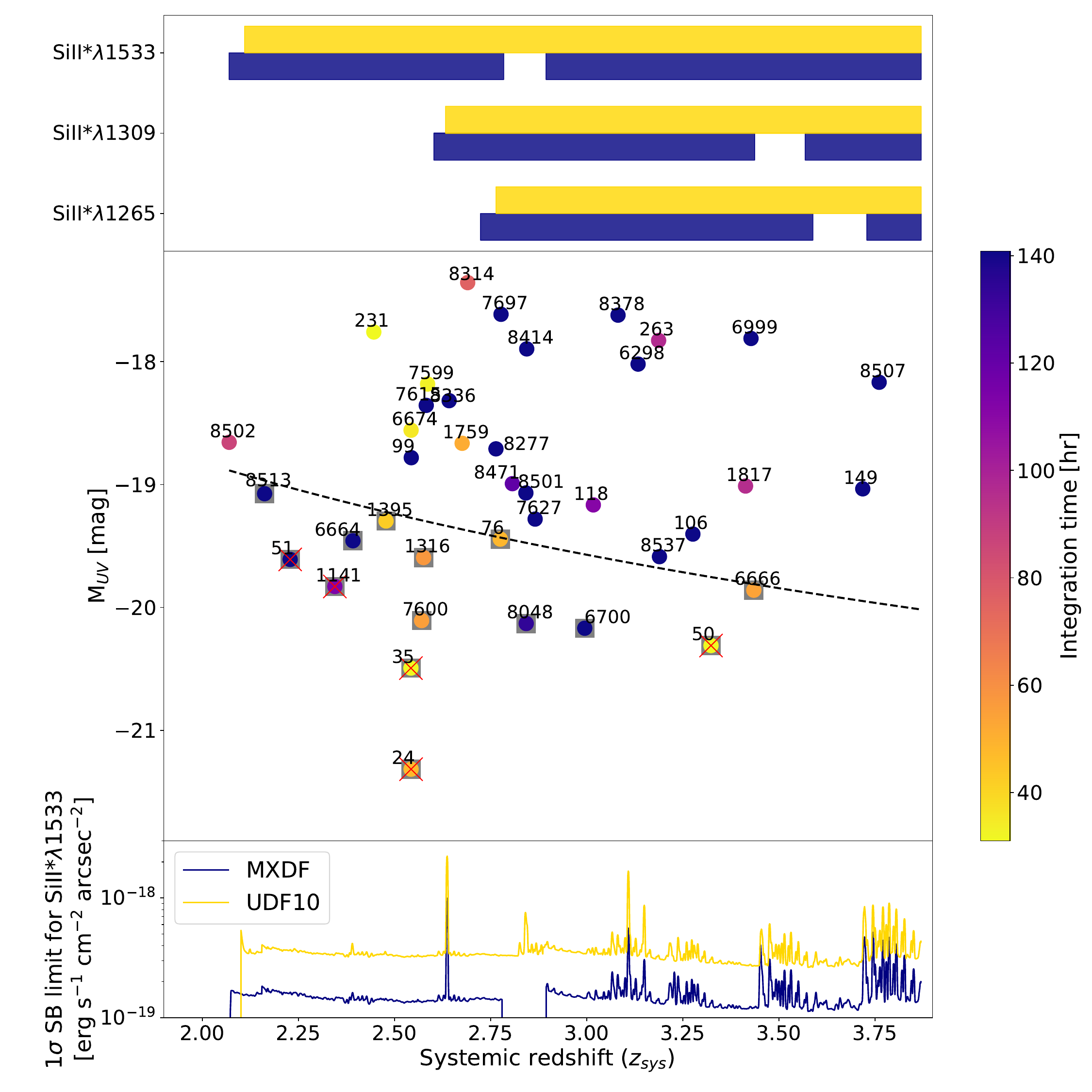}
      \caption{ Redshift distribution of our sample. (Top) Redshift ranges covered for the \SiIIf\ $\lambda$1265, 1309, 1533 lines. Navy and yellow shades indicate \mxdf\ and \udft, respectively. The gaps for the navy bars are the AO gap (see Section \ref{subsec:data}). (Middle) The $M_{\rm UV}$ and $z_{\rm sys}$ distribution of our sample with MUSE IDs. The color bar indicates the integration time. The stacked subsample is enclosed by gray squares. The $M_{\rm UV}$ corresponding to $m_{\rm UV}=26.0$ (magnitude cut for the stacking sample) at each redshift is shown by the black dashed line. The individually-detected \SiIIf\ halos are indicated by red crosses (see Section \ref{subsec:res_nb_test} ). (Bottom) 1$\sigma$ SB limits for \SiIIf\ $\lambda$1533 in \mxdf\ (navy) and \udft\ (yellow), which are converted from the median surface brightness limits on the sensitivity table in \citet{bacon_muse_2023} using the narrow band widths for \SiIIf\ $\lambda$1533 for given redshifts (see Section \ref{subsec:res_nb_sb}). The \SiIIf\ halos tend to have bright $M_{\rm UV}$. }
         \label{fig:MUV_zsys}
\end{figure*}

We construct a sample of galaxies at $z=2.07$--$3.87$, for which the [C\,{\sc iii}]$\lambda$1907, C\,{\sc iii}]$\lambda$1909 doublet nebular lines and at least one of the \SiIIf\ $\lambda$1265, 1309, 1533 lines are redshifted into the MUSE wavelength range. The available redshift ranges for the three \SiIIf\ lines are shown in the top panel in Fig. \ref{fig:MUV_zsys}. We require sources to have a secure spectroscopic redshift with $ZCONF\geq2$ and a C\,{\sc iii}] signal-to-noise ratio (S/N) of $S/N>3$. We visually inspect C\,{\sc iii}] in spectra and images extracted from MUSE cubes to ensure these criteria. We use systemic redshifts measured by the C\,{\sc iii}] line rather than reference redshifts (REFZs) in the DR2 catalog, which are given by $\lya$ at $z\gtrsim3$ or absorption lines at $z\simeq2$. \SiIIf\ lines are expected to be located at the systemic redshifts \citep[e.g.,][]{france_diffuse_2010,dessauges-zavadsky_rest-frame_2010,jaskot_linking_2014,wang_systematic_2020}, though redshifted \SiIIf\ lines are also reported, which could be due to the contamination of the absorption at the blue edge for low-resolution spectra or stacking of spectra for different sources \citep[e.g., ][]{shapley_rest-frame_2003, erb_physical_2010,berry_stacked_2012}. As \SiIIf\ lines are extremely faint, secure measurements of systemic redshifts enable us to not rely on the presence of \SiIIf\ in 1D spectra on a galaxy scale and to directly search for spatially extended \SiIIf\ in narrowband images extracted from MUSE cubes. We limit our sample to galaxies with a secure and isolated HST counterpart \citep[based on][MAG\_SRC$=$RAF]{rafelski_uvudf_2015} with a high HST matching confidence level of MCONF$\geq3$ in the DR2 catalog. In total, we have 39 galaxies, which are summarized in Table \ref{table:sample}. Absolute UV magnitudes ($M_{\rm UV}$) and systemic redshifts ($z_{\rm sys}$) are shown in the middle panel of Fig. \ref{fig:MUV_zsys}. The $M_{\rm UV}$ is calculated at rest-frame 1600 ~\AA\ by fitting 2 or 3 HST bands with a power-low model (with parameters being an amplitude and a UV slope $\beta$). Figure \ref{fig:MUV_zsys} also indicates the redshift coverage of the three \SiIIf\ lines (top panel) and the surface brightness limits for each field (bottom panel). The stellar masses and star formation rates (SFRs) of the sample are shown in Fig. \ref{fig:sfr_ms} in Appendix.

We would like to note that our sample is constructed simply with the deepest data for this pilot study. We discuss possible strategies to extend the sample in Section \ref{subsec:discussion_extension}. 

\subsection{Stacked sample}\label{subsec:sample_stack} 
The subsample used for the stacking analysis is restricted to bright UV-continuum galaxies using an apparent magnitude cut of $m_{\rm UV}=26.0$,
in order to enhance the S/N of the stacked images (see Section \ref{subsec:method_stack} and Fig. \ref{fig:MUV_zsys}). The origin of \SiIIf\ is predicted to be continuum pumping as mentioned in Section \ref{sec:intro}, and a bright UV continuum is necessary to have a bright \SiIIf\ emission line. Indeed, \SiIIf\ emission lines are stronger in 1D stacked spectra for subsamples of MUSE LAEs with brighter UV continuum (such as those with higher stellar masses and with brighter UV magnitudes) than the subsample counterparts in \citet{feltre_muse_2020}. The number of sources for the stacking subsample is 14. We would like to note that applying fainter magnitude cuts indeed reduces the S/Ns of the \SiIIf\ lines.

\subsection{Masks}\label{subsec:sample_mask} 
In order to exclude pixels that might be affected by bright neighboring objects in \SiIIf\ narrowband images extracted from the minicubes, we create neighboring object masks. First, we define target pixels. We use an HST segmentation map and mask pixels not corresponding to a main target on a 5''$\times$5'' HST/F775W cutout for each object. It is convolved with the MUSE moffat PSF at the redshifted \SiIIf\ wavelength for the main target and rebinned to match the MUSE pixel scale. Then, we apply a threshold value of 0.1 to the peak-normalized convolved cutout to delimit the spaxels of the target. Lower threshold values such as 0.05 cause contamination of neighboring objects inside the target spaxels. Using a higher threshold value of 0.2 does not change the main results in this paper but enhances the contaminated pixels outside the following neighboring object masks. Therefore, we adopt the value 0.1. 

Second, we use a 5''$\times$5'' HST/F775W cutout whose sky and the main target are masked with the HST segmentation map for each source. This cutout is convolved with the MUSE PSF and rebinned to match the MUSE pixel scale. Then, we define pixels brighter than a certain threshold on the cutout, except for the target pixels, as masked pixels. The applied threshold value is half of the value above in the absolute sense (without normalization).  

\section{Results}\label{sec:result}
\subsection{Extracted images and surface brightness profiles}\label{subsec:res_nb_sb} 
For each source, we create a \SiIIf\ emission narrowband (NB) image from the continuum-subtracted minicube. We sum fluxes in a window of $\pm$200 km s$^{-1}$ around the \SiIIf\ wavelength, excluding NaN (masked) spaxels, wavelength slices affected by OH skylines, and the AO gap in the cube. The variance image of the NB is created from that of the cube with error propagation. A UV continuum broadband (BB) is created from the median filtered original minicube with a window of $\pm$300 wavelength slices (375~\AA) around the \SiIIf\ wavelength (masking below $+400$~km~s$^{-1}$ from the $\lya$ wavelength) to obtain a good S/N. We also exclude NaN in the cube and the AO gap. If the wavelength window is not fully covered, we use as many spectral slices as possible. The variance image of the BB is assumed to be identical to that of the mean-filtered BB created from the same cube with error propagation. We apply a neighboring object mask (see Section \ref{subsec:sample_mask}) and measure SB profiles of these two bands using \textsf{PHOTUTILS} \citep{bradley_astropyphotutils_2021}. The aperture and annuli are centered at the HST counterpart position with 1-pixel ($0\farcs2$) annuli from $0\farcs1$ to $0\farcs9$ and then 2-pixel annuli at $1\farcs2$ and $1\farcs6$, followed by a 3.5-pixel annulus at $2\farcs15$, to enhance the S/N at outer radii. The azimuthally-averaged radial SB profiles are calculated from the effective areas and the measured flux above. We normalize the SB profile of the UV continuum at the innermost radius (R=$0\farcs1$) to that of \SiIIf.

\begin{figure*}
   \centering
\includegraphics[width=0.99\linewidth]{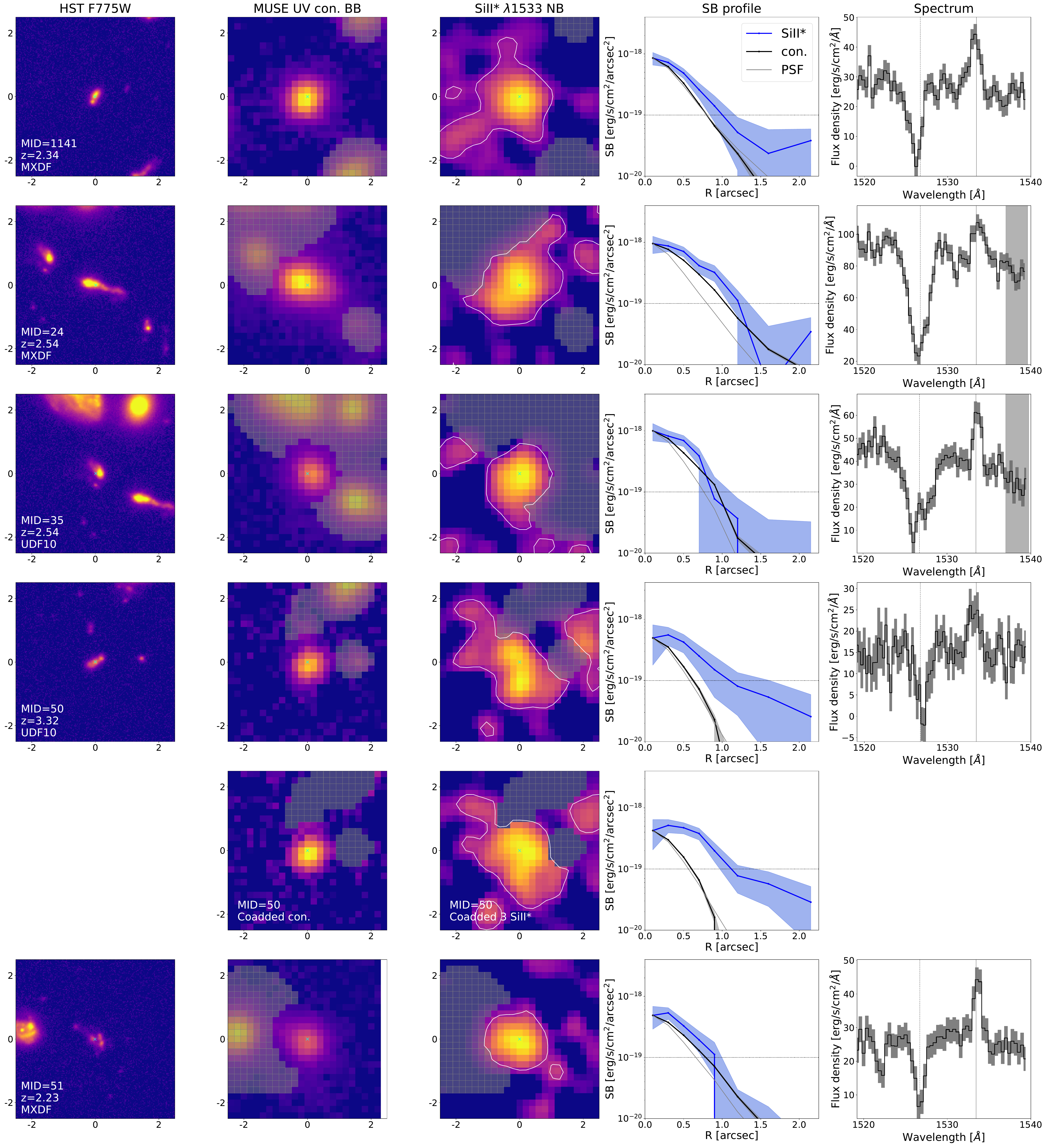}
      \caption{(From left to right) The HST image,  MUSE UV-continuum BB image, MUSE \SiIIf\ NB image, SB profile, and \SiIIf\ spectrum for the five sources with \SiIIf\ halo detections. Each row except the fifth shows the \SiIIf\ $\lambda1533$ line for a different object. The fifth row shows a coadded result of three \SiIIf\ lines for MID=50 at $z=3.32$, whose three \SiIIf\ lines (\SiIIf\ $\lambda$1265, \SiIIf\ $\lambda$1309, and \SiIIf\ $\lambda$1533) are covered with MUSE. The image size is five arcsec. Gray shades on the images indicate masked spaxels. For presentation purposes, NB images are smoothed with a Gaussian kernel with $\sigma=1.5$~spaxel. White contours correspond to $1\times10^{-19}$ \sbl\ \citep[the typical 3$\sigma$ SB limit in MXDF for an unresolved emission,][]{bacon_muse_2023}. Cyan crosses on images indicate the HST center. The SB profiles of the UV continuum shown by black lines are normalized at the innermost radius (R=$0\farcs1$) to the SB profiles of \SiIIf\ shown by blue lines. Blue and gray shades represent 1$\sigma$ uncertainties of the SB for \SiIIf\ emission and UV continuum, respectively. Gray solid and dotted lines show the SB profile of the PSF normalized at the center and the typical 3$\sigma$ SB limit of the NBs in MXDF for unresolved emission, respectively. The rest-frame ``reference spectrum'' and its 1$\sigma$ uncertainty are shown by a black solid and a gray shade, respectively. The reference spectrum for each object is chosen from those extracted by five different methods to have high-S/N in \citet{bacon_muse_2023}. The wavelengths for \SiIIf\ emission and \SiII\ absorption lines are indicated by gray solid and dotted lines, respectively. MID$=24$ and MID$=35$ have broad C\,{\sc iv} $\lambda$$\lambda$1548, 1550 lines, which are masked by gray shades at the red edge of the spectrum. The MUSE \SiIIf\ NB images show a clear extension of \SiIIf\ compared to the UV continuum on the MUSE BB images, which can be confirmed with their SB profiles.  
              }
         \label{fig:5sourcesSB}
\end{figure*}

The top panels in Fig. \ref{fig:5sourcesSB} show the \SiIIf\ emission NB, the UV-continuum BB, and the SB profile for a highlighted object, MID$=1141$. The details of this object are explained later (Section \ref{subsec:res_nb_test}).

\subsection{Identification of \SiIIf\ halos}\label{subsec:res_nb_test} 

To identify extended \SiIIf\ emission, we compare the shape of the azimuthally-averaged radial SB profiles of the UV continuum and the \SiIIf\ emission. We take account of uncertainties in SB profiles of \SiIIf\ emission only (blue shades in Fig. \ref{fig:5sourcesSB}) as those of UV continuum are negligibly small (gray shades). If more than two adjacent data points of the \SiIIf\ SB profiles deviate by more than $1 \sigma$ from the normalized continuum, we identify it as extended \SiIIf. The significance of the presence of the halo with this criterion corresponds to more than 97.5\%\footnote{The probability for two adjacent data points with $>1 \sigma$, which is caused by chance due to noises (i.e., fake detections), is $\bigl(\frac{1-0.6826}{2}\bigr)^2$=2.5\% with an assumption that noises follow the normal distribution.}. We would like to note that, as we use circular photometry, our test may miss non-circular halos as discussed in \citet{kusakabe_muse_2022} for Ly$\alpha$ halos. Developing and testing non-circular methods requires high-S/N objects for reliable assessments, which our sample is not eligible for. Our halo search is not complete in that sense but secure. 

Among the 39 sources, we detect five \SiIIf\ halos for the \SiIIf\ $\lambda$ 1533 line (MID$=24$, $35$, $50$, $51$, $1141$; see Fig. \ref{fig:5sourcesSB}). These are the first detections of individual \SiIIf\ halos. The \SiIIf\ lines in the DR2 reference spectra show that there are no significant velocity shifts for the line peaks of the \SiIIf\ $\lambda1533$ (Fig. \ref{fig:5sourcesSB}; the systemic redshifts are given by the [C\,{\sc iii}]\ $\lambda$1907, C\,{\sc iii}]\ $\lambda$1909 doublet nebular lines, see Section \ref{subsec:sample}). Meanwhile, peaks of the \SiII\ $\lambda1527$ absorption lines show velocity offsets, which have often been observed in the down-the-barrel analysis and interpreted as gas flows \citep[e.g.,][]{du_redshift_2018}\footnote{\SiII\ $\lambda1527$ line is a commonly used tracer of outflows seen along the line of sight. Due to the resonant scattering involved in the 1527 transition (see Fig. \ref{fig:energy_levels}), observed \SiIIf\ $\lambda1533$ has information on the other line of sight directions. Therefore, these two lines may exhibit different velocity offsets, even if the origin of \SiIIf\ $\lambda1533$ is continuum pumping influenced by resonant scattering.}.

Among the five \SiIIf\ halo sources, MID=$1141$, shown in the top panels of Fig. \ref{fig:5sourcesSB}, is the best-case object. The \SiIIf\ emission profile shows three data points that deviate by more than $1\sigma$ from the normalized continuum. This object is well-isolated and without a spatial offset of \SiIIf\ from the UV continuum. The images are not significantly affected by masks, which makes the SB profile test secure. The \SiIIf\ emission line can be clearly confirmed on the spectrum shown in Fig. \ref{fig:5sourcesSB}. MID$=24$ and MID=$35$ are a pair of galaxies, which could contaminate each other's 5'' $\times$ 5'' images despite the neighboring object mask. Moreover, MID=$35$ is an X-ray detected AGN \citep{luo_chandra_2017}, whose \SiIIf\ halo might have a different origin from other sources. The \SiIIf\ NB of MID=$50$ shows a peak offset from the UV center measured with the HST data, with two neighboring LAEs in the DR2 catalog, which spatially overlap with the extended \SiIIf\ emission. The \SiIIf\ SB of MID=$51$ is noisy even though it meets our halo criterion. Considering these details on the five sources, we highlight MID=$1141$ in the following sections. 

As shown in Fig. \ref{fig:MUV_zsys}, the five individual halos tend to have bright UV magnitudes, which is qualitatively consistent with the continuum pumping scenario of \SiIIf\ emission. We also discuss possible differences in SFRs and $M_\star$ between \SiIIf\ $\lambda$1533 halo sources and nondetected sources in Fig. \ref{fig:sfr_ms}, and also examine the dependence of detections of halos on 1$\sigma$ SB limits for \SiIIf\ NBs in Fig. \ref{fig:muv_sb}, although they are beyond the scope of this paper to conclude them. Overall, we cannot conclude them due to the small sample size, but UV magnitudes and redshifts of the sources might be essential.

The redshifts of the individually-detected \SiIIf\ $\lambda$1533 halos range from 2.23 to 3.32 with four of them at $z_{\rm sys}\lesssim2.5$. The nondetections at $z>3.5$ would be caused by the lack of UV-bright sources (see Figs. \ref{fig:MUV_zsys} and \ref{fig:muv_sb}) if the origin of \SiIIf\ is continuum pumping. It could also be due to the high noise levels of MUSE data at red wavelengths \citep[severe night sky and low sensitivity; ][]{bacon_muse_2023} and the cosmic dimming effect on the SB profiles (by a factor of 2.7 between $z=2.5$ and $z=3.5$). In case the circumgalactic \SiIIf is intrinsically fainter at $z\sim2$ than at $z>3.5$, it implies a rapid evolution of metal enrichment in the CGM, which is qualitatively consistent with the redshift evolution of the number density of \MgII\ absorbers \citep{matejek_survey_2012,zhu_jhu-sdss_2013} as well as the prediction from a chemical evolution model with UniverseMachine (\citealt{behroozi_most_2018}; Nishigaki et al. in prep.).

Another interesting result is that we cannot confirm extended emission for any of the other \SiIIf\ lines ($\lambda$1265, 1309). In the optically thin limit, the strengths of fluorescent lines are expected to depend on the strength of the paired absorption lines, which are determined by their oscillator strengths, $f$. In particular, the \SiII\ $\lambda1260$ resonant absorption line corresponding to the \SiIIf\ $\lambda$1265 line has a ten times larger $f$ than that of the \SiII\ $\lambda1527$ resonant absorption line corresponding to the \SiIIf\ $\lambda$1533. However, in the simulations of \citet{mauerhofer_uv_2021} there is no significant difference in the equivalent widths of \SiIIf\ $\lambda$1265 and \SiIIf\ $\lambda$1533 absorption lines because both lines are saturated (Mauerhofer et al. in prep.). In such an optically thick regime, the equivalent widths for the corresponding fluorescent emission (\SiIIf\ $\lambda$1265 and \SiIIf\ $\lambda$1533) are similar in the simulations. Therefore, non detections of \SiIIf\ $\lambda$1265 as well as \SiIIf\ $\lambda$1309 would not be due to fainter total fluxes for these lines than for the 1533 line. The reason for nondetections of \SiIIf\ $\lambda$1265 and \SiIIf\ $\lambda$1309 halos are not clear. The samples for \SiIIf\ $\lambda$1265, 1309 lines are limited to high redshifts ($z_{\rm sys}>2.8$ and $>2.6$, respectively), and none of them are bright in apparent UV magnitudes, which are directly related to S/Ns (see Fig. \ref{fig:MUV_zsys}). Bright sources would be crucial for the detection of diffuse \SiIIf\ lines if the origin of \SiIIf\ is continuum pumping. However, we cannot distinguish whether it is caused by differences in lines, samples, or noise levels. We note that the \SiIIf\ $\lambda$ 1533 lines are also brighter than the other two lines in the stacked MUSE LAE spectra of \citet{feltre_muse_2020} and \citet{kramarenko_linking_2024}, which might also be due to the difference in objects stacked for different \SiIIf\ lines.

One of the individually-detected objects, MID=$50$ at $z_{\rm sys}=3.32$, has 3 \SiIIf\ lines covered with MUSE. We coadd the NBs and BBs for the 3 \SiIIf\ lines by taking the median (with the same method as used in Section \ref{subsec:method_stack}) and show them in the fifth row of Fig. \ref{fig:5sourcesSB}. Extended \SiIIf\ is confirmed with a better S/N, which implies the \SiIIf\ $\lambda$1265, 1309 lines are also spatially extended.

\begin{figure*}
   \centering
   \includegraphics[width=\hsize]{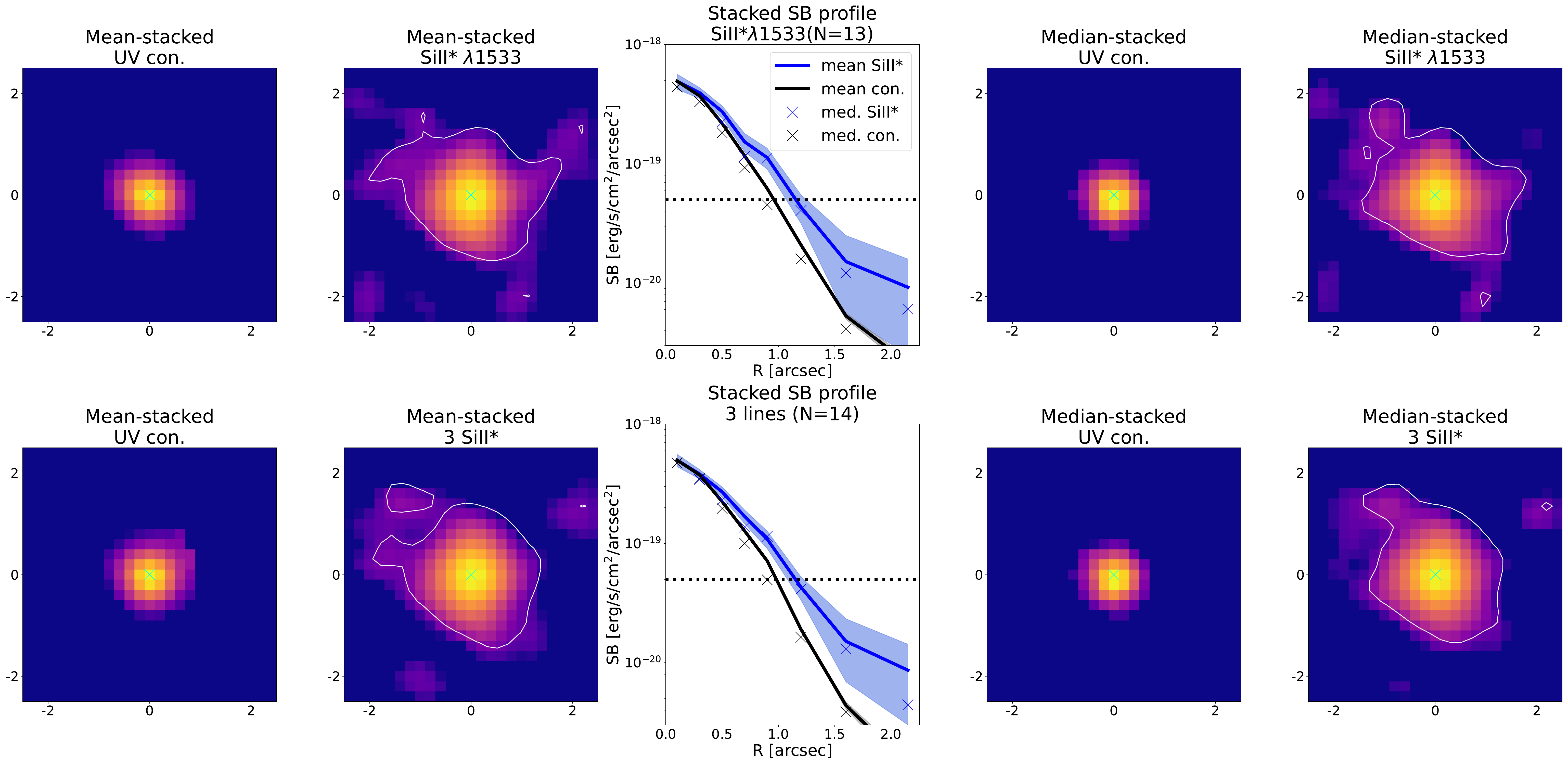}%
      \caption{Stacking results. From left to right, we show the mean-stacked UV continuum image, the mean-stacked \SiIIf\ emission image (5 arcsec each), the mean surface brightness profiles, the median-stacked UV continuum image, and the median-stacked \SiIIf\ emission image. Top and bottom panels represent stacking results of the \SiIIf\ $\lambda$1533 line (N=$13$) and those of all \SiIIf\ lines  (22 images for 14 sources), respectively. Cyan crosses on images indicate the HST center. White contours correspond to $5\times10^{-20}$ \sbl\, which is also shown by black dotted lines on the SB profiles. The SB profiles of the mean-stacked \SiIIf\ and UV continuum are indicated by the blue and black lines, respectively, while the median-stacked \SiIIf\ and UV continuum are shown by the blue and the black crosses, respectively. The blue and gray shades represent 1$\sigma$ uncertainties of the SB for \SiIIf\ emission and UV continuum, respectively. Following Fig. \ref{fig:5sourcesSB}, SB profiles for the continuum are normalized at the innermost radius (R=$0\farcs1$) to those for \SiIIf\ emission. The mean and median stacked \SiIIf\ NB images show a clear extension of \SiIIf\  compared to the UV continuum on the stacked MUSE BB images, which can be confirmed with their SB profiles.
              }
         \label{fig:stackedSB}
\end{figure*}

\subsection{Stacking analysis}\label{subsec:method_stack} 
To test the general presence of \SiIIf\ halos, we stack 14 UV-bright sources (see Section \ref{subsec:sample_stack}). The numbers of sources with the magnitude cut of $m_{\rm UV}=26.0$ available for each of the \SiIIf\ $\lambda$1265, 1309, 1533 lines are five, four, and 13, respectively, which are limited by the MUSE wavelength coverage and the AO gap (see Fig. \ref{fig:MUV_zsys})\footnote{The MIDs for the \SiIIf\ $\lambda$1265 subsample are 50, 76, 6666, 6700, and 8048. Those for the \SiIIf\ $\lambda$1309 are 50, 76, 6700, and 8048. Those for the \SiIIf\ $\lambda$1533 are 24, 35, 50, 51, 76, 1141, 1316, 1395, 6664, 6666, 6700, 7600, and 8513.}. Therefore, we only stack a subsample for the \SiIIf\ $\lambda$1533 line and take the mean and median of the \SiIIf\ NBs and BBs without weighting, while masking neighboring objects. We also stack 14 sources by combining the NBs and BBs for the three lines. The variance images for the mean stacking, which we use for our halo test, are created based on error propagation. Then, we apply the SB profile test for the mean-stacked images with the same method as used in Sections \ref{subsec:res_nb_sb} and \ref{subsec:res_nb_test}, as the uncertainties can be calculated correctly for the mean but not for the median. 

The results of this stacking experiment are shown in Fig. \ref{fig:stackedSB}. As shown in the top row, we confirm the existence of a \SiIIf\ $\lambda$1533 halo with five data points that deviate by more than $1 \sigma$ from the normalized continuum profile. The median profile of \SiIIf\ $\lambda$1533 is consistent with the mean profile and deviates from the median continuum as well, in particular at large radii. These tests may suggest that the detected halo in the mean profile is not dominated by a few extreme objects. This is the first secure detection of a stacked \SiIIf\ halo. It may imply that the individually detected sources are not extreme cases, \SiIIf\ halos must be common for UV-bright galaxies at $z=2$--$4$.

The bottom row of Fig. \ref{fig:stackedSB} shows the results of stacking all the \SiIIf\ narrowbands and corresponding UV continuum bands for the 14 sources (three lines; in total 22 images). We find that the mean-stacked \SiIIf\ deviates by more than $1\sigma$ from the mean-stacked UV continuum for six data points. The median-stacked \SiIIf\ is consistent with the mean-stacked \SiIIf, and the deviation from the normalized continuum is similarly seen with the median-stacked profiles.

We note that removing MID$=35$, which is an X-ray detected AGN with extended \SiIIf\ (see Section \ref{subsec:res_nb_test}), does not change the results and that the other 13 sources do not have a counterpart in the deep X-ray catalog \citep{luo_chandra_2017}.
We also note that stacking without individually-detected \SiIIf\ halos results in nondetections of extended \SiIIf. As shown in Fig. \ref{fig:MUV_zsys}, all the individually-detected sources have a bright $M_{\rm UV}$ (and also a bright $m_{\rm UV}$) and are located at relatively low redshifts. It is not surprising that removing such five sources results in nondetections. If the origin of the \SiIIf\ is continuum pumping, as we expect, it then should be more challenging to detect \SiIIf\ halos without these bright sources. We also stack various subsamples including all the sources without individually-detected halos and get nondetections of extended \SiIIf\ emission. Future instruments such as BlueMUSE \citep{richard_bluemuse_2019} or a larger sample compiling MUSE and KCWI archive data will allow us to investigate the dependence of \SiIIf\ halos on sample properties such as $M_{\rm UV}$ (see Section \ref{subsec:discussion_extension}). It will enable us to draw a firm conclusion of whether \SiIIf\ halos are indeed common for UV-bright galaxies at $z=2$--$4$ or not.  

\section{Discussion}\label{sec:discussion}
\subsection{Testing photon conservation}\label{subsec:discussion_1141}
\citet{shapley_rest-frame_2003} suggest that the origin of the \SiIIf\ is continuum pumping (see Section \ref{sec:intro} for other origins). For the \SiIIf\ $\lambda1533$  line, Si$^+$ in neutral clouds makes a \SiII\ $\lambda1527$ resonant absorption line, and then a fraction of Si$^+$ deexcites to the fine-structure level, emitting \SiII\ fluorescence photons at 1533~\AA\ (see Fig. \ref{fig:energy_levels}). As the \SiII\ is a resonant line, the \SiIIf\ line can be regarded as an escape channel of photons from resonant scattering. When dust opacity is high in the ISM and the CGM, dust attenuation for \SiIIf\ can be greater than that for the UV continuum, depending on the number of scatterings. if additional dust attenuation of \SiIIf\ is negligible and the escape is isotropic, then the equivalent width (EW) of \SiIIf\ emission and that of the corresponding resonant absorption should be comparable because of photon conservation.

However, it is possible for the absorption and emission to have different EWs for one given direction of observations even if the origin is continuum pumping without dust, because of the complex gas distribution in front of stars in galaxies \citep[e.g.,][]{prochaska_probing_2011,carr_semi-analytical_2018}. Simulations predict that absorption lines depend sensitively on the direction of observations \citep{mauerhofer_uv_2021,gazagnes_interpreting_2023}. Below, we assume an isotropic case.

\begin{figure*}
   \centering
   \includegraphics[width=0.95\linewidth]{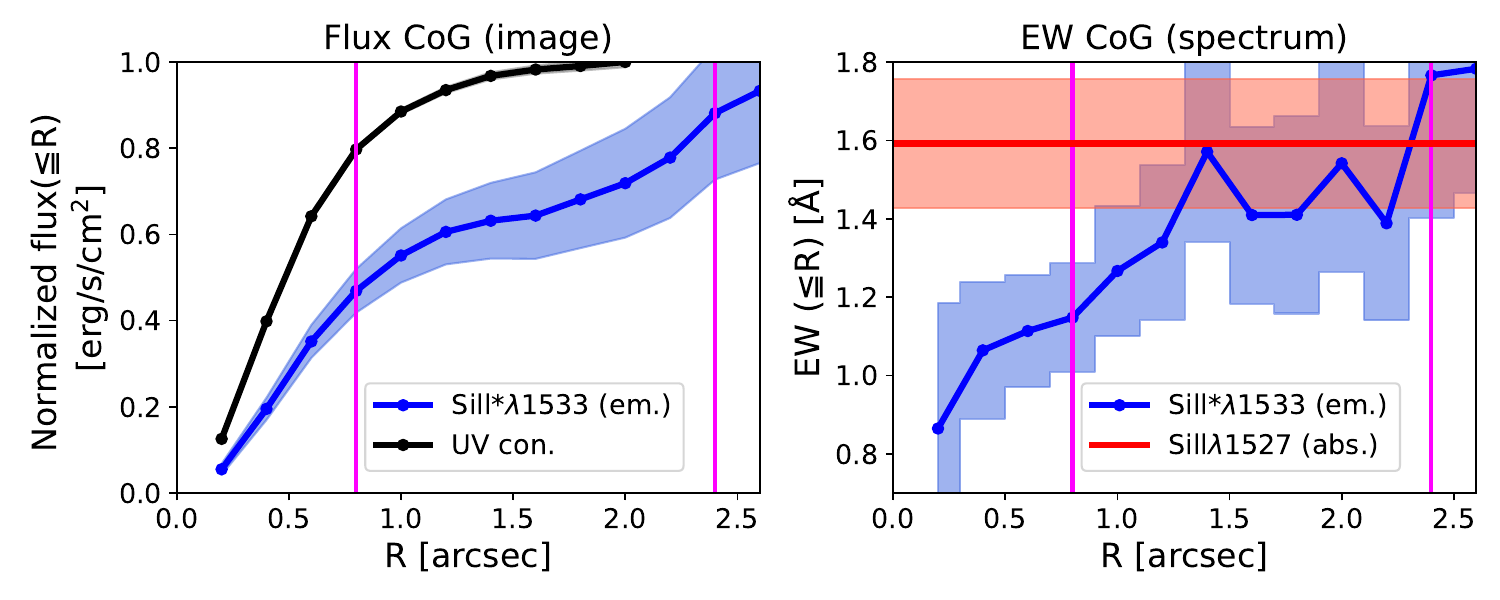} 
      \caption{  Curve of growth (CoG) for \SiIIf\ flux and EW for MID$=1141$. Left: Normalized fluxes within radius R as a function of R measured from the \SiIIf\ NB image (blue line) and the UV continuum BB image (black line). The blue shade indicates $1\sigma$ uncertainties for the \SiIIf\ flux. The vertical magenta lines show R=$0\farcs8$ (galaxy scale) and R=$2\farcs4$ (CGM scale), respectively. Right: EW(\SiIIf) CoG as a function of R shown by the blue line with the blue shade indicating $1\sigma$ uncertainties. EWs are measured in spectra extracted from the original minicube with growing apertures around the HST center. The red line and shade show the EW for \SiII\ absorption at R=$0\farcs8$ and its $1\sigma$ uncertainty, respectively. The left panel shows that the galaxy-scale aperture includes about 80\% of the continuum but only 50\% of the \SiIIf, while the CGM-scale aperture captures most of the \SiIIf\ flux. The right panel shows that the absorption EW on the galaxy scale and the fluorescent emission EW on the CGM scale are consistent with each other (photon conserved), which implies that the origin of extended \SiIIf\ emission is the continuum pumping.
              }
         \label{fig:1141_cog}
\end{figure*}

To investigate the scenario of continuum pumping, we use the isolated, high-SN object MID$=1141$ (see Section \ref{subsec:res_nb_test}) and test photon conservation between \SiII\ absorption and associated \SiIIf\ emission, which is commonly assessed with the EWs of these lines \citep[e.g.,][]{prochaska_simple_2011,wang_systematic_2020}. First, we check the flux curve of growth (CoG) for the continuum and the emission line to determine the spatial extent. Second, we measure the EW curve of growth for \SiIIf\ emission and compare it with the absorption EW. We measure EWs from the MUSE data (without HST), which makes the EW comparison fair. Ideally, we should measure the curve of growth for the absorption. However, due to the limited depth and spatial resolution of our data, we have to assume that the absorption follows the same spatial profile as that of the continuum. The left panel of Fig. \ref{fig:1141_cog} shows normalized fluxes within growing apertures for continuum (black) and \SiIIf\ $\lambda1533$ (blue). Fluxes are measured from the \SiIIf\ $\lambda1533$ NB image and the corresponding UV continuum BB image (Section \ref{subsec:res_nb_sb}) with \textsf{PHOTUTILS} and the neighboring object mask with corrections of masked areas. On a galaxy-scale of R$=0\farcs8$, the aperture includes about 80\% of the continuum but only 50\% of the \SiIIf. An R$=2\farcs4$ aperture captures most of the \SiIIf\ flux. 

\begin{figure*}
\centering
   \includegraphics[width=0.95\linewidth]{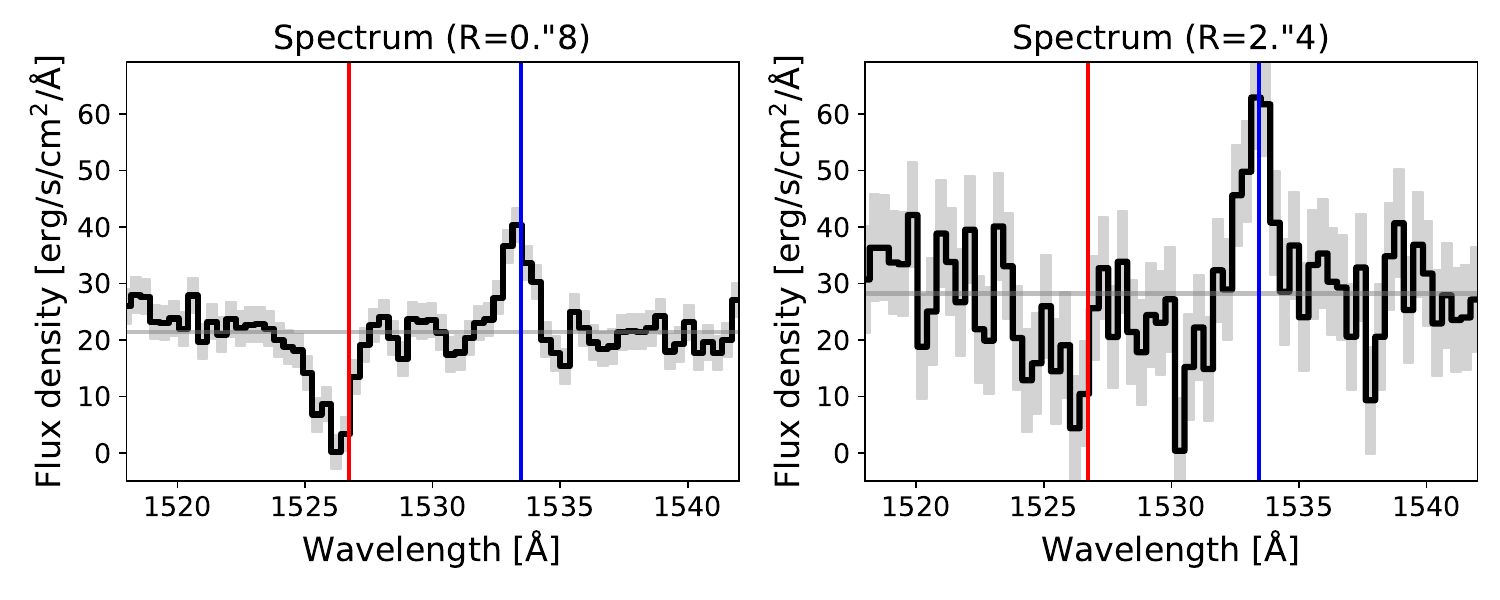}
      \caption{  Galaxy-scale and CGM-scale spectra for MID=1141. Left: The black line shows the spectrum extracted from the original minicube with a $0\farcs8$ aperture (radius) around the HST center. The gray solid line indicates the continuum. The gray shade shows 1$\sigma$ uncertainties. The vertical red and blue lines indicate the rest-frame wavelength of \SiII\ absorption and \SiIIf\ emission (EW$_{\rm abs}$(\SiII\ $\lambda1527$)$=1.6\pm0.2$~\AA\ and EW$_{\rm em}$(\SiIIf\ $\lambda1533$)$=1.1\pm0.1$~\AA). Right: The spectrum extracted from a $2\farcs4$ aperture around the HST center (EW$_{\rm em}$(\SiIIf\ $\lambda1533$)$=1.8\pm0.4$~\AA). The spectra indicate that the photon conservation works on the CGM scales.
              }
         \label{fig:1141_spec}
\end{figure*}

The right panel shows rest-frame EWs for \SiIIf\ $\lambda1533$ emission within growing apertures (blue), compared with the EW for \SiII\ $\lambda1527$ absorption measured at R$\leq0\farcs8$, EW$_{\rm abs}$(\SiII\ $\lambda1527$)$=1.6\pm0.2$~\AA\ (red). Rest-frame EWs are measured in spectra extracted from the original minicube (to which the neighboring object mask is applied) with growing apertures. The continuum flux density is measured by the median filtering each spectrum from 50-slice shorter wavelength (62.5\AA) than the absorption to 50-slice longer wavelength than the emission. Absorption and emission fluxes are measured as differences from the continuum. The EW for \SiIIf\ $\lambda1533$ is EW$_{\rm em}$(\SiIIf\ $\lambda1533$)$=1.1\pm0.1$~\AA\  at R$\leq0\farcs8$, which is smaller than EW$_{\rm abs}$(\SiII\ $\lambda1527$)$=1.6\pm0.2$~\AA. EW$_{\rm em}$(\SiIIf\ $\lambda1533$) increases with R and reaches EW$_{\rm em}$(\SiIIf\ $\lambda1533$)$=1.8\pm0.4$~\AA\ at R$\leq2\farcs4$, which is consistent with EW$_{\rm abs}$(\SiII\ $\lambda1527$) at R$\leq0\farcs8$ within the 1$\sigma$ uncertainties (see Fig. \ref{fig:1141_spec} for the spectra), although the EWs have large measurement errors. Therefore, the absorption EW on the galaxy scale and the fluorescent emission EW on the CGM scale are consistent with each other. This is compatible with photon conservation on the CGM scale, albeit with large uncertainties. It implies that the origin of extended \SiIIf\ emission can be the continuum pumping as predicted under the assumption of isotropic escape and no dust. In that case, either a small amount of dust, a low optical depth, or a combination of both are implied. Unfortunately, data for the other four sources with individual \SiIIf\ halos are either noisy or have accompanying galaxies inside their 5''$\times$5'' minicubes, which prevents us from obtaining a firm conclusion about the mechanisms for the four sources. The trend of increasing EW of \SiIIf\ emission with radius beyond the galaxy scale is predicted by simulations \citep{mauerhofer_uv_2021}, which are compared with slit spectroscopy of low-redshift galaxies in \citet[][see also \citealt{wang_systematic_2020}]{gazagnes_interpreting_2023}.

\subsection{Comparisons with zoom simulations}\label{subsec:discussion_simu}

We compare our stacked \SiIIf\ $\lambda1533$ profile with those from the cosmological zoom-in simulations of \citet{mauerhofer_uv_2021}, which are used in \citet{gazagnes_interpreting_2023} and \citet{blaizot_simulating_2023}. \citet{mauerhofer_uv_2021} present three snapshots containing a relatively small galaxy (such as MUSE LAEs) at $z=3.0$, $3.1$, and $3.2$. The stellar mass for the simulated galaxy is $M_\star \sim2\times10^9$~M$_\odot$, which is similar to the median value of those for the 14 galaxies used in the stacking analysis, $M_\star\sim3\times10^9$~M$_\odot$ (ranging $M_\star \sim1\times10^{9}$--$2\times10^{10}$~M$_\odot$,  \citealt{bacon_muse_2023}, see Fig. \ref{fig:sfr_ms}). The SFRs of the simulated galaxy for the three snapshots are $2.2$~M$_\odot$~yr$^{-1}$, $4.2$~M$_\odot$~yr$^{-1}$, and $5.0$~M$_\odot$~yr$^{-1}$, which are similar to the median value of the observed galaxies, 3.4 M$_\odot$~yr$^{-1}$ (ranging $SFR=1.5$--$24.5$ M$_\odot$~yr$^{-1}$, \citealt{bacon_muse_2023}, see Fig. \ref{fig:sfr_ms}). \SiIIf photons in this simulations originate from continuum pumping. We provide mock MUSE cubes for observations from 12 directions for each snapshot in the rest frame. This original mock cube has a size of $300\times300\times110$ pixels corresponding to $7\farcs5\times7\farcs5\times16.6$~\AA\ (from 1521.83~\AA\ to 1538.42~\AA) with a spaxel scale of $0\farcs025$/pix. Then, to match the size of observed and simulated galaxies, we measure the median size of the simulated galaxy among the 36 directions. A UV-continuum BB image is created from each median-filtered original cube with a window from 1528.5~\AA\ to 1531.5~\AA\ (the continuum between \SiII\ absorption and \SiIIf\ emission). We define the galaxy center by identifying the brightest pixel, as in \citet{gazagnes_interpreting_2023}, for each UV continuum BB image after HST PSF convolution \citep[F606W in][]{rafelski_uvudf_2015}. The half-light radius (R$_{50}$) is measured around the galaxy center with the original mock cube (without HST PSF convolution) using \textsf{PHOTUTILS}. The median R$_{50}$ for the simulated galaxy is $0\farcs0875$ (3.5 pix). This is 1.7 times smaller than the median R$_{50}$ for the 14 stacked galaxies (R$_{50}$=$0\farcs15$ with F606W after PSF correction). Since we do not have a simulated galaxy that is large enough in \citet{mauerhofer_uv_2021}, we manually rescale the original mock cube by a factor of 1.7 by interpreting the spaxel scale as $0\farcs0425$/pix, which is referred to as an intrinsic mock cube below. The intrinsic mock cubes are cut out around the galaxy center defined above and are 5 arcsec each. The 5'' cubes are convolved with the MUSE PSF and line spread function (LSF) described in \citet{bacon_muse_2023} at the median redshift of the stacked sample ($z=2.68$). Then the convolved mock cubes are rebinned to match the MUSE cubes in the rest frame. We apply similar analyses to the 36 mock MUSE cubes as used for the observational data (see Section \ref{sec:result}). A slight difference from the method for the MUSE data is continuum subtraction. Continuum cubes used for continuum subtraction are created using median filtering from 1528.5~\AA\ to 1531.5~\AA\, which are also used for the UV-continuum BB images. The 1$\sigma$ uncertainties for stacked mock data are derived from 15.87 to 84.12 percentiles of SB profiles among individual mocks. We would like to note that the simulations assume cosmological parameters of $H_0 = 67.11$ km s$^{-1}$ Mpc$^{-1}$ and $\Omega_{\rm m} = 0.3175$, which should not have a significant impact on the results below. We convert distances in the unit of arcsec for the simulated data to those in the unit of kpc at $z=2.68$ with the {\it Planck} 2018 cosmological model \citep{planck_collaboration_planck_2020} as the galaxy sizes are matched and MUSE PSF are convolved in the unit of arcsec.

\begin{figure*}
   \centering
   \includegraphics[width=\hsize]{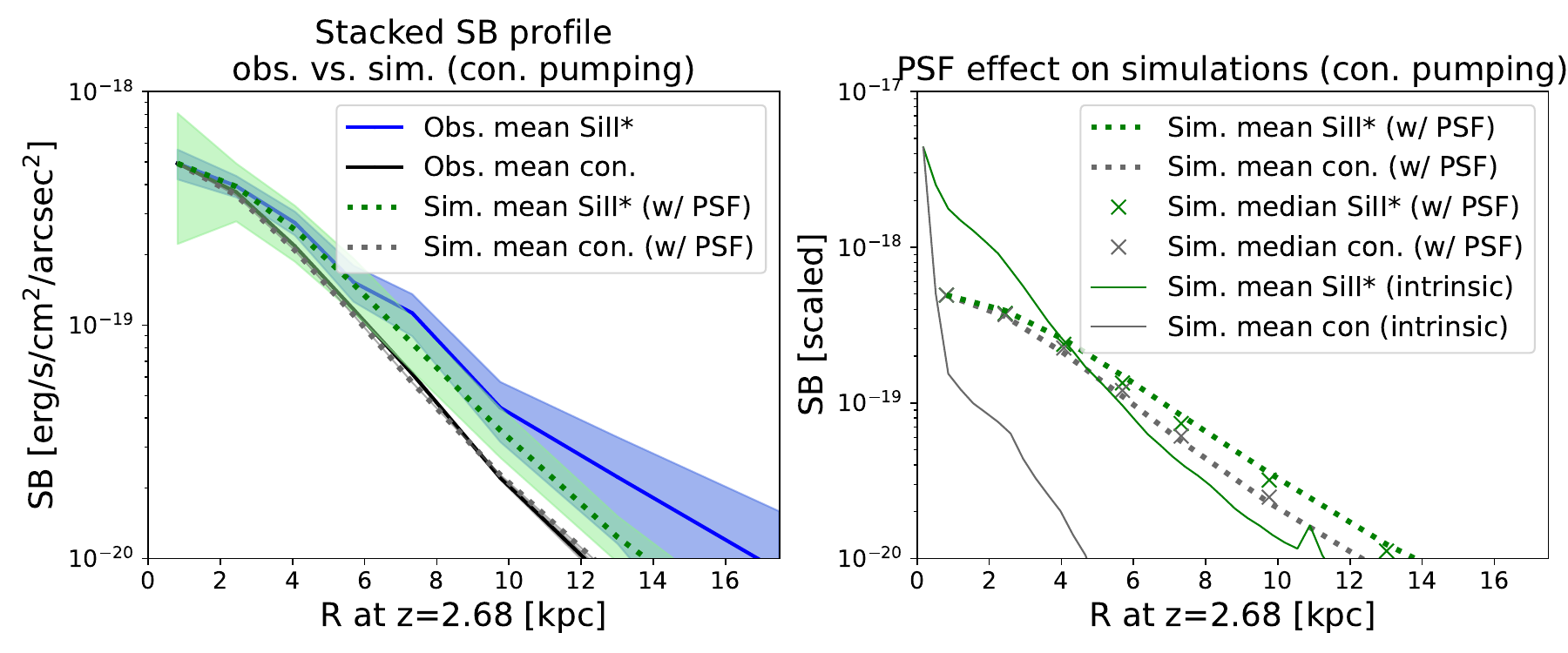}
      \caption{  Stacked SB profiles of our observations and of zoom-in simulations from \citet{mauerhofer_uv_2021}. Left: Comparison of stacked SB profiles. The blue and black solid lines show observed mean-stacked SB profiles of \SiIIf\ $\lambda1533$ and continuum, respectively. The blue shade indicates 1$\sigma$ uncertainties for \SiIIf\ $\lambda1533$. The observed continuum is normalized at $R=0$ to the peak of the observed \SiIIf\ $\lambda1533$. The green and gray dotted lines (shades) show mean-stacked SB profiles of \SiIIf\ $\lambda1533$ and continuum, respectively, for the simulations after PSF and LSF convolution (their 1$\sigma$ uncertainties, which correspond to 15.87 to 84.13 percentile of SB profiles for individual mocks). They are normalized at $R=0$ to the peak of the observed \SiIIf\ profile. Right: Stacked SB profiles for simulations. The green and gray dotted (solid) lines show the mean-stacked SB profiles of \SiIIf\ $\lambda1533$ and continuum, respectively, with (without) PSF and LSF convolution. The green and gray crosses show the median-stacked SB profiles of \SiIIf\ $\lambda1533$ and continuum, respectively, with PSF and LSF convolution. They are scaled by the same factor as used in the left panel. The left panel shows that observed stacked SB profiles can be reproduced within $1\sigma$ uncertainties by the simulations, which account for continuum pumping. 
              }
         \label{fig:comp_simu}
\end{figure*}

The left panel of Fig. \ref{fig:comp_simu} shows a comparison of the observed and simulated mean-stacked SB profiles. The simulated UV continuum (gray dotted line) is well matched with the observations (black solid line), thanks to our manual 1.7 times expansion. The simulated \SiIIf\ $\lambda1533$ profile (green dotted line) is more extended than the simulated UV continuum (gray dotted line) and consistent with the observed \SiIIf\ $\lambda1533$ (blue solid line), though simulated \SiIIf\ halo is slightly less extended than observed within $1\sigma$ uncertainties.

The mean profiles can be biased toward bright halo sources, so we check whether the mean-stacked profiles are consistent with the median-stacked profiles for the \SiIIf\ and the continuum (green and gray crosses, respectively) in the right panel of Fig. \ref{fig:comp_simu}. We confirm that they are consistent with each other, as for the MUSE observations of \SiIIf\ $\lambda1533$. The right panel also compares simulated mean-stacked SB profiles with PSF and LSF convolution versus without convolution, indicated by dotted and solid lines, respectively. The simulated galaxy indeed has significantly more extended \SiIIf\ $\lambda1533$ than the UV continuum in the intrinsic mock cubes, but the difference is mostly hidden by the MUSE PSF convolution\footnote{We check individual mocks and find that the shapes of the individual SB profiles (green dotted line) are fully dominated by the PSF.}.

\begin{figure*}
   \centering
   \includegraphics[width=\hsize]{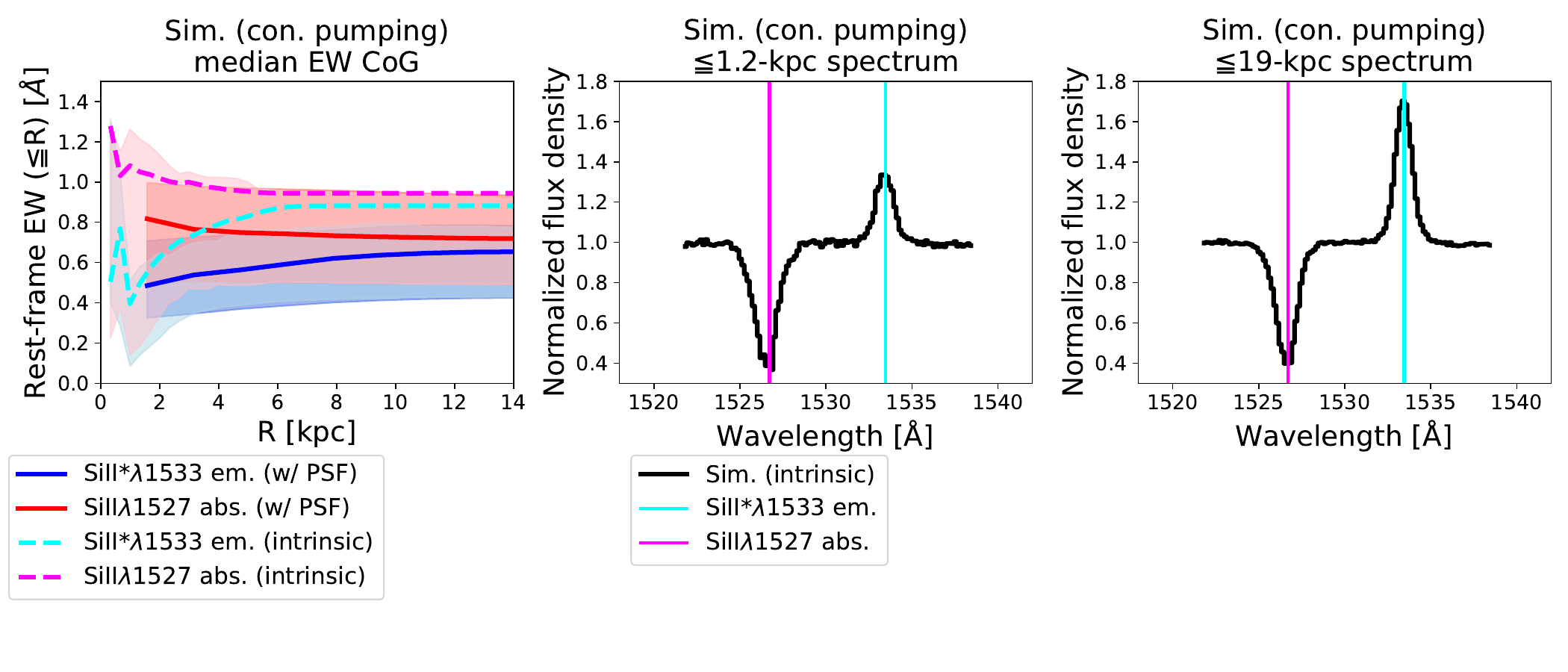}
      \caption{  EW CoG and spectra of the simulations with continuum pumping. Left: The median cumulative EW in rest-frame as a function or R. The blue and red solid lines (shades) indicate the EW of \SiIIf\ $\lambda1533$ emission and \SiII\ $\lambda1527$ absorption lines (their 1$\sigma$ uncertainties) with MUSE PSF convolution, respectively. The cyan and magenta dashed lines (shades) show the intrinsic EW of \SiIIf\ $\lambda1533$ emission and \SiII\ $\lambda1527$ absorption lines (their 1$\sigma$ uncertainties). Middle: The median spectrum extracted from the intrinsic mock cube with a 1.2 kpc aperture (galaxy scale) shown by the black line. The cyan and magenta lines show the wavelength of \SiIIf\ $\lambda1533$ emission and \SiII\ $\lambda1527$ absorption lines, respectively. Right:  The median spectrum extracted from the intrinsic mock cube with a 19 kpc aperture (CGM scale). The EWs of \SiIIf\ $\lambda1533$ emission and \SiII\ $\lambda1527$ absorption are consistent at a CGM scale for both intrinsic cubes and PSF-convolved cubes, which are also seen on the mock spectra.
              }
         \label{fig:simu_ew_spec}
\end{figure*}

We also check EW CoG of \SiIIf\ $\lambda1533$ emission and \SiII\ $\lambda1527$ absorption lines for the simulations as shown in the left panel of Fig. \ref{fig:simu_ew_spec}. The EWs of \SiIIf\ $\lambda1533$ emission and \SiII\ $\lambda1527$ absorption are consistent at a CGM scale for both intrinsic cubes and PSF-convolved cubes, which are also seen on the mock spectra (see the middle and right panels of Fig. \ref{fig:simu_ew_spec}). It means that the photon conservation works for the simulations with continuum pumping in the sense that the scattered photons are not more attenuated by dust than that stellar continuum. This is consistent with our observations for MID=1141. 

From these comparisons, we conclude that our simulations with continuum pumping scenario can reproduce the observations. This lends extra support for the interpretation of \SiIIf\ emission as a signature of continuum pumping. We note that the selection bias with the UV magnitude cut may have an impact on the observed result and this comparison. The mock observations are re-scaled by a factor of $1.7$ to correct for the fact that the simulated galaxy is smaller than the observed ones (see above). The simulated galaxy is fainter than the observed one, though the range of absolute magnitudes overlaps ($M_{\rm UV}$ from -18 to -19). It would be interesting to test the selection bias with a larger and brighter simulated galaxy, or several, maybe at a slightly lower redshift.

\subsection{Possible extension of the sample}\label{subsec:discussion_extension}

We discuss possible strategies to extend the sample with existing data in this section. There are a few ways to increase the sample size: (1) sources at $z>3.87$, which have other nebular line detections such as the O\,{\sc iii}]\ $\lambda\lambda$1661, 1666 doublet and He\,{\sc ii}\ $\lambda1640$ in the same catalog and the same field (with the same integration time), (2) sources located in shallower fields such as \mosaic\ field (9$\times$9-pointing with 10-hour integration, in the same catalog of \citealt{bacon_muse_2023}) and MUSE-WIDE field \citep[1-hour integration][]{herenz_muse-wide_2017, urrutia_muse-wide_2019, kerutt_equivalent_2022}, (3) UV-bright sources at the same redshifts in other fields with similarly long integration times ($>30$ hours) in MUSE archival data \citep[e.g.,][]{fossati_muse_2019,lofthouse_muse_2020}. The samples (1) and (2) are tested with the MUSE DR2 catalog (not including MUSE WIDE) in the same manner. We find that these less restrictive selection criteria do not help to increase the number of detections of \SiIIf\ halos. Therefore, (3) using deep MUSE archival data for UV-bright galaxies with the same selection criteria would be the best way to extend the sample in the future.

In addition, (4) gravitationally lensed sources observed with MUSE \citep[e.g.,][Claeyssens et al. in prep.]{richard_atlas_2021,claeyssens_lensed_2022}, would be useful to investigate more compact \SiIIf\ halos in the future (see the right panel of Fig. \ref{fig:comp_simu}), which would be hidden by the PSF without the power of the magnification. We would like to note that they are not included in this pilot study as magnification can add complications and uncertainties. To investigate \SiIIf\ halos at higher redshifts ($z\sim3$--$6$), whose \SiIIf\ lines are covered with MUSE, we could use (5) sources with systemic redshifts measured by nebular lines with JWST, and (6) MUSE LAEs, whose systemic redshifts can be estimated by empirical relationships based on the Ly$\alpha$ line \citep{verhamme_recovering_2018}. We do not include these sources in this pilot study, considering uncertainties in wavelength calibration \citep[e.g.,][]{maseda_jwstnirspec_2023, deugenio_jades_2024,meyer_jwst_2024} and more severe cosmic dimming effects. It is future work for us to investigate higher-$z$ sources.

\section{Conclusions}\label{sec:conclusions} 
To study the spatial distribution of the metal-enriched cool CGM at cosmic noon in emission, we search for extended \SiIIf\ emission (fluorescent lines) using the MUSE HUDF data and catalog with 30-140 hour integration times. We construct a sample of 39 galaxies with systemic redshifts at $z=2.07$--$3.87$, at which the [C\,{\sc iii}]$\lambda$1907, C\,{\sc iii}]$\lambda$1909 doublet nebular lines and at least one of the \SiIIf\ $\lambda$1265, 1309, 1533 lines are redshifted into the MUSE wavelength range. Our major results are summarized as follows. 

\begin{enumerate}
\item Five individual \SiIIf\ $\lambda1533$ halos are confirmed to be present from statistical tests with surface brightness profiles. These are the first detections of extended \SiIIf\ emission. 

\item We stack images of \SiIIf\ $\lambda$1533 line for a subsample of 13 UV-bright galaxies. We confirm the presence of stacked \SiIIf\ $\lambda 1533$ halos, which may imply that metal-enriched CGM is common for UV-bright galaxies. We also stack images of \SiIIf\ $\lambda$1265, 1309, 1533 for 14 UV-bright galaxies and detect a \SiIIf\ halo. If we stack fainter galaxies or remove the five individually-detected halos from stacking, we get nondetections of \SiIIf\ halos. We need a larger sample to draw a firm conclusion.

\item We find that the EW of the absorption line and fluorescent emission line are roughly equal when integrated out to large distances for MID=1141. This suggests that photons are conserved and that the emission line is mostly due to pumping from the stellar continuum.

\item We test the presence of \SiIIf\ halos in zoom-in simulations, which account only for continuum pumping. After re-scaling the mock observations to correct for the fact that the simulated galaxy is smaller than the observed one, we find that the simulated halos are consistent with the observed stacked halo within $1\sigma$ uncertainties. This trends extra support for the interpretation of \SiIIf\ emission as a signature of continuum pumping.

\item The best way to extend our sample with existing data in the future is using deep MUSE and KCWI archival data for UV-bright galaxies with the same selection criteria.

\end{enumerate}
KCWI or BlueMUSE \citep{richard_bluemuse_2019} will allow us to investigate HI Ly$\alpha$ halos for \SiIIf\ halos detected with MUSE. BlueMUSE will also make it possible to individually detect \SiIIf\ halos of more diverse sources and to study their statistical properties thanks to a more moderate cosmic dimming effect at $z\simeq2$ (by a factor of three, compared to $z\simeq3$). Such larger samples would be useful to improve the stacking experiments (see also Section \ref{subsec:discussion_extension}) and to study spectral variations at different radii from the centers of galaxies as done for Ly$\alpha$  \citep{wisotzki_nearly_2018, guo_median_2024, guo_spatially-resolved_2023}. Such studies will be advanced more with better sensitivities and higher spatial resolutios achieved by extremely large telescopes such as Thirty-meter telescope (TMT), one of whose key science cases is CGM mapping with WFOS \citep{skidmore_thirty_2015}, as well as European-Extremely large telescope (E-ELT).

\begin{acknowledgements}
We thank the anonymous referee for constructive comments and suggestions. We would like to express our gratitude to Charlotte Simmonds for useful comments and suggestions. HK acknowledges support from the Japan Society for the Promotion of Science (JSPS) Overseas Research Fellowship (202160056) as well as JSPS Research Fellowships for Young Scientists (202300224, 23KJ2148). HK thanks Yuri Nagai, an academic assistant at NAOJ, for her wonderful support. VM acknowledges support from the Nederlandse Organisatie voor Wetenschappelijk Onderzoek (NWO) grant 016.VIDI.189.162 ('ODIN'). AV and TG are supported by the SNF grant PP00P2 211023. T.N. acknowledges support from Australian Research Council Laureate Fellowship FL180100060. I.P. acknowledges funding by the European Research Council through ERC-AdG SPECMAP-CGM, GA 101020943. This work is based on observations taken by VLT, which is operated by European Southern Observatory. This research made use of Astropy\footnote{\url{http://www.astropy.org}}, which is a community-developed core Python package for Astronomy \citep{the_astropy_collaboration_astropy_2013, the_astropy_collaboration_astropy_2018}, and other software and packages:\textsf{MPDAF} \citep{piqueras_mpdaf_2019}, \textsf{PHOTUTILS}, \textsf{Numpy} \citep{harris_array_2020}, \textsf{Scipy} \citep{virtanen_scipy_2020}, and \textsf{matplotlib} \citep{hunter_matplotlib_2007}. 

\end{acknowledgements}

\begin{appendix}
\section{Overview of the sample}\label{ap:sample}

\begin{figure}[h]
   \includegraphics[width=\hsize]{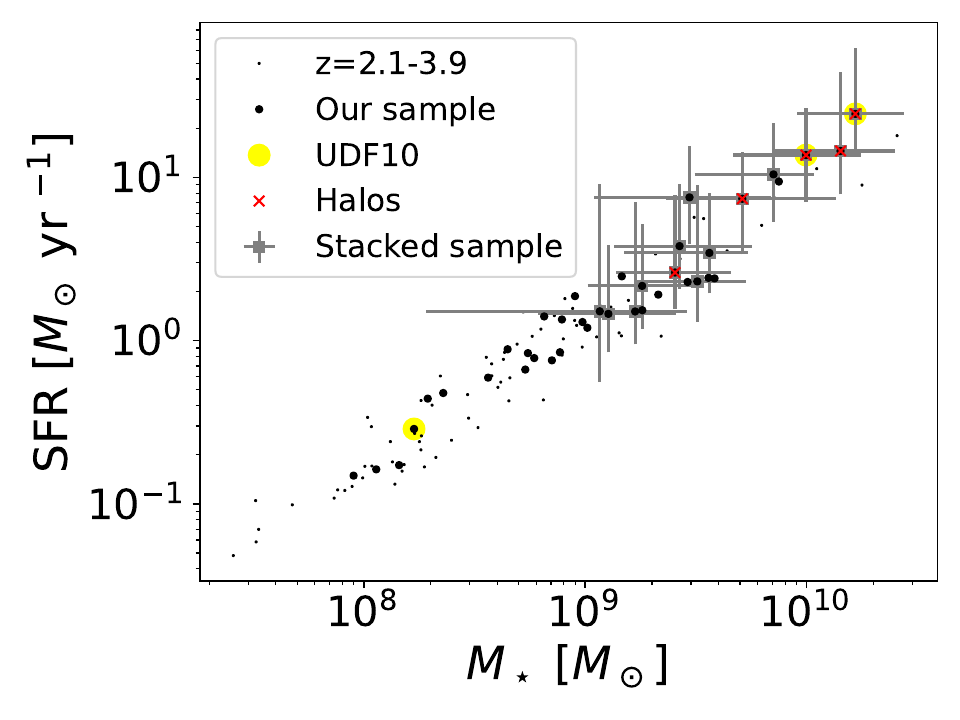}
      \caption{SFR and  $M_\star$ for the full sample (black circles), the stacked sample (gray squares with error bars), the three sources in \udft\ (yellow circles), and the individual \SiIIf\ halos (red crosses), compared to those of galaxies $z=2.1$--$3.9$ in the DR2 catalog (gray dots). Due to the large 1$\sigma$ uncertainties in SFR and $M_\star$ \citep{bacon_muse_2023}, error bars are shown only for the stacked sample" for clarity and conciseness. The full sample lies along the relation for the galaxies at $z=2.1$--$3.9$ in the same catalog, which suggests that our sample is not biased towards starburst galaxies. 
     }
         \label{fig:sfr_ms}
\end{figure}

The basic information about our sample is summarized in \ref{table:sample}. Among the 39 sources, only three sources are selected from \udft\ with an integration time of 31 hours (ID$=35^*$, $50$, and $231$), and the remaining 36 are from \mxdf\ with integration times ranging from 32 hours to 141 hours. The small number of \udft\ sources is due to the overlaps between \mxdf\ field with \udft\ \citep[see figure 2 in][]{bacon_muse_2023}. The \mxdf\ data are assigned to most of the sources observed both in \mxdf\ and \udft\ because of higher qualities (except for ID$=35^*$ in our sample). 

Figure \ref{fig:sfr_ms} shows the distribution of the SFRs and the $M_\star$ of our sample, which are derived from spectral energy distribution (SED) fitting in the DR2 catalog \citep[using Prospector, see section 6.4 more details][]{bacon_muse_2023}. Although the 1$\sigma$ uncertainties in SFR and $M_\star$ are large, the 39 sources (black circles) lie along the relation for galaxies at $z=2.1$--$3.9$ in the same catalog (gray dots, star-formation main sequence). This suggests that our sample is not biased toward starburst galaxies. The 14 sources used in the stacking analysis (black circles without gray squares with error bars) tend to have higher SFR and $M_\star$ than the other sources (black circles without enclosed by gray squares), which are expected from the selection criteria with UV magnitudes for the stacked sample. 

Next, we examine potential differences between the two fields. Sources in \mxdf\ are widely distributed and include low-mass sources with faint magnitudes (see also Fig. \ref{fig:MUV_zsys}). Among the three sources in \udft\ (yellow circles), two are located at the massive end of the sequence in Fig. \ref{fig:sfr_ms}, while the remaining one has a low stellar mass. As expected from the depth of the MUSE data, the fractions of low-mass sources with low SFRs and faint magnitudes in \mxdf\ are higher than those in \udft, regardless of thresholds. However, 
because of the small sample size in \udft\ ($N=3$), our main results are unlikely to be significantly affected by the depth of \udft. Studying the fractions and extents of halos is beyond the scope of this paper, which requires a larger sample and careful assessment of the inhomogeneous data set (see Appendix \ref{ap:udf10_mxdf} for the examination of the SB limits for our sources). In Appendix \ref{ap:properties}, we discuss potential differences in SFR and $M_\star$ for individually-detected halos and non-halos.

\begin{table*}[h]
\caption{Overview of the sample}\label{table:sample}
\begin{tabular}{llllllll}
\hline 
  ID &        RA &        DEC &  $z_{\rm sys}$ &       $M_{\rm UV}$ [mag] &  Integration [h] & Stacked & Individual Halo \\
\hline
  24$^{+}$ & 53.160088 & -27.776356 &  2.543301 & 24.0 &               47 &   Yes &  Yes \\
  35$^{+, *}$ & 53.160587 & -27.776120 &  2.542948 & 24.8 &               31 &   Yes &  Yes \\
  50 & 53.162849 & -27.771626 &  3.323586 & 25.4 &               31 &   Yes &  Yes \\
  51 & 53.165183 & -27.781614 &  2.228735 & 25.4 &              140 &   Yes &  Yes \\
  76 & 53.157832 & -27.777562 &  2.775538 & 26.0 &               48 &   Yes &   No \\
  99 & 53.163077 & -27.785605 &  2.543808 & 26.5 &              140 &    No &   No \\
 106 & 53.163726 & -27.779076 &  3.276452 & 26.3 &              140 &    No &   No \\
 118 & 53.157088 & -27.780269 &  3.017429 & 26.4 &              110 &    No &   No \\
 149 & 53.167866 & -27.778617 &  3.718381 & 26.9 &              140 &    No &   No \\
 231 & 53.162101 & -27.772563 &  2.446336 & 27.4 &               31 &    No &   No \\
 263 & 53.162688 & -27.776638 &    3.1873 & 27.9 &               96 &    No &   No \\
1141 & 53.165074 & -27.793661 &  2.344985 & 25.3 &              113 &   Yes &  Yes \\
1316 & 53.169306 & -27.793799 &  2.575916 & 25.7 &               57 &   Yes &   No \\
1395 & 53.174057 & -27.790421 &  2.478122 & 25.9 &               42 &   Yes &   No \\
1759 & 53.170789 & -27.793294 &  2.676148 & 26.7 &               51 &    No &   No \\
1817 & 53.156111 & -27.789788 &  3.413466 & 26.8 &               95 &    No &   No \\
6298 & 53.169249 & -27.781255 &  3.133906 & 27.6 &              140 &    No &   No \\
6664 & 53.162345 & -27.784443 &  2.392062 & 25.7 &              140 &   Yes &   No \\
6666 & 53.159576 & -27.776719 &   3.43494 & 26.0 &               53 &   Yes &   No \\
6674 & 53.166556 & -27.775259 &  2.543293 & 26.7 &               34 &    No &   No \\
6700 & 53.168281 & -27.781056 &  2.995106 & 25.4 &              140 &   Yes &   No \\
6999 & 53.159601 & -27.791271 &  3.427612 & 28.0 &              140 &    No &   No \\
7599 & 53.154801 & -27.780390 &  2.586264 & 27.1 &               32 &    No &   No \\
7600 & 53.153765 & -27.784885 &  2.570747 & 25.2 &               55 &   Yes &   No \\
7615 & 53.170752 & -27.785994 &  2.582759 & 26.9 &              140 &    No &   No \\
7627 & 53.170800 & -27.783148 &   2.86605 & 26.2 &              140 &    No &   No \\
7697 & 53.169279 & -27.781394 &  2.777275 & 27.8 &              140 &    No &   No \\
8048 & 53.167939 & -27.792722 &  2.842856 & 25.3 &              133 &   Yes &   No \\
8277 & 53.167383 & -27.790112 &  2.764135 & 26.7 &              140 &    No &   No \\
8314 & 53.174925 & -27.785724 &   2.69058 & 28.0 &               75 &    No &   No \\
8336 & 53.157657 & -27.783316 &  2.642334 & 27.0 &              140 &    No &   No \\
8378 & 53.159007 & -27.781481 &  3.081696 & 28.0 &              140 &    No &   No \\
8414 & 53.171088 & -27.789138 &  2.844166 & 27.6 &              140 &    No &   No \\
8471 & 53.169458 & -27.778191 &  2.806879 & 26.5 &              121 &    No &   No \\
8501 & 53.167806 & -27.791894 &  2.841994 & 26.4 &              140 &    No &   No \\
8502 & 53.174309 & -27.787684 &   2.07006 & 26.2 &               86 &    No &   No \\
8507 & 53.162236 & -27.783618 &  3.760994 & 27.8 &              140 &    No &   No \\
8513 & 53.158786 & -27.784608 &  2.162213 & 25.9 &              140 &   Yes &   No \\
8537 & 53.169650 & -27.788081 &  3.189507 & 26.1 &              140 &    No &   No \\
\hline
\end{tabular}
\tablefoot{ID: MUSE ID, RA: right ascension, DEC: declination, $z_{\rm sys}$: systemic redshift measured with C\,{\sc iii}] doublet (see Section \ref{subsec:sample}), $M_{\rm UV}$: Absolute UV magnitudes (see Section \ref{subsec:sample}),  Integration: integration time, Stacked: The galaxy is stacked or not (see Section \ref{subsec:sample_stack})?, Individual Halo: \SiIIf\ halo is individually detected or not (see Section \ref{subsec:sample_stack}) $^{+}$: ID=24 and ID=35 are pair galaxies.  $^*$: ID=35 is an AGN (see Section \ref{subsec:res_nb_test}). It is observed for 45 hours in \mxdf, but \udft\ data are adopted in \citet{bacon_muse_2023}. } 
\end{table*}
\FloatBarrier

\section{Potential effects of different depths and sample properties}\label{ap:diff}
\subsection{Potential effects of different depths of the MUSE data for our sample}\label{ap:udf10_mxdf}

\begin{figure*}[t]
   \includegraphics[width=\linewidth]{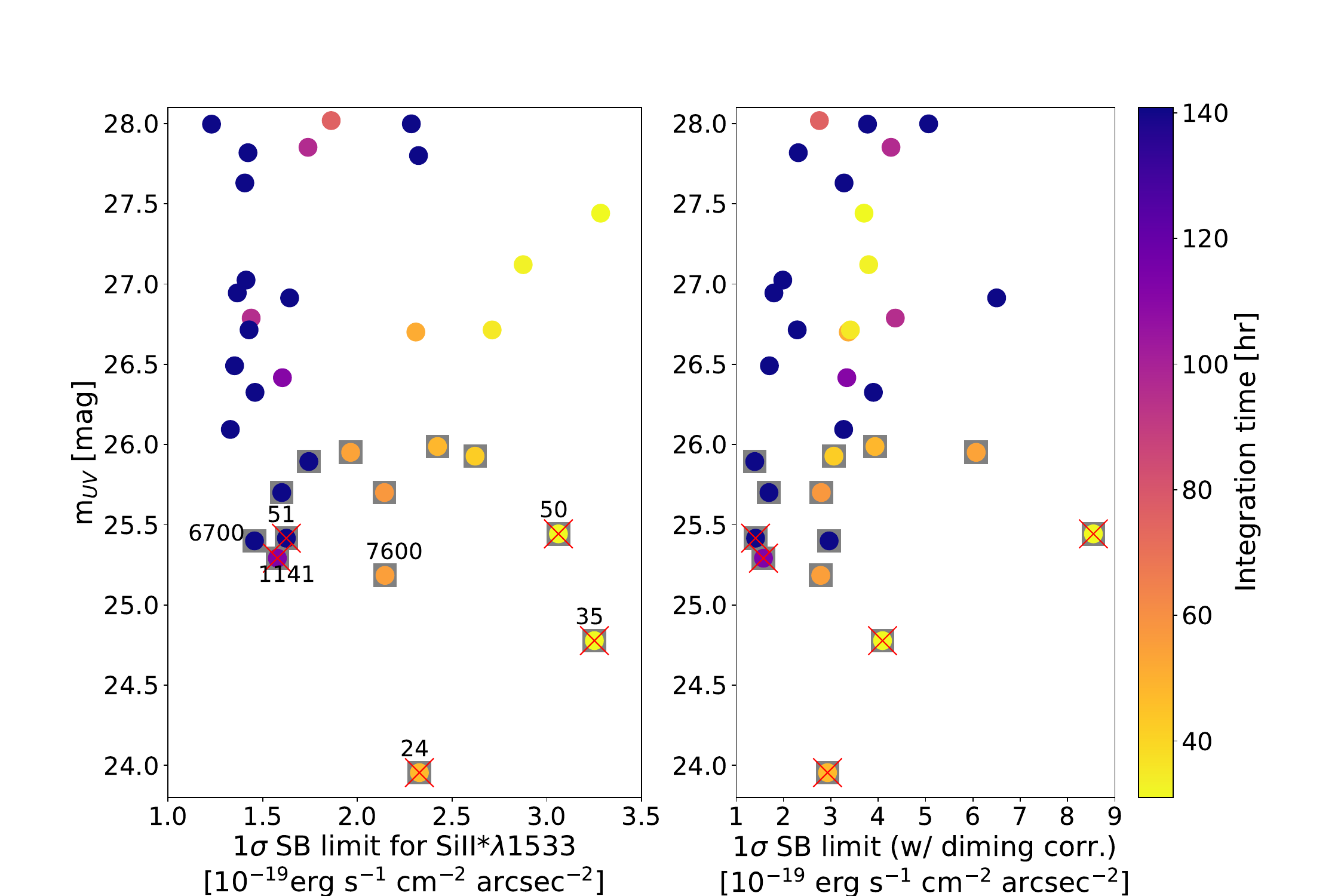}
      \caption{Distributions of 1$\sigma$ SB limits of \SiIIf\ $\lambda$1533 and apparent UV magnitudes for our sample. The left panel shows 1$\sigma$ SB limits on a 1-arcsec$^2$ scale for the \SiIIf\ $\lambda$1533 NBs, while the right panel shows 1$\sigma$ SB limits multiplied by (1+$z$)$^4$/(1+$z_{1141}$)$^4$ to account for the effect of cosmic dimming in a relative manner. The color bar indicates the integration time. The stacked subsample is enclosed by gray squares, and the individually detected \SiIIf\ halos are indicated by red crosses (see Section \ref{subsec:res_nb_test} ). The MUSE IDs for the seven objects with $m_{\rm UV}\lesssim25.5$, which are brighter than the faintest individually-detected halo (ID$=51$) are shown in the left panel. Detection rates of halos are expected to depend on the SB limits of the data set, but the panels do not show any clear trend, likely due to the limited sample size and different sample properties.
     }
         \label{fig:muv_sb}
\end{figure*}

As mentioned in Section \ref{subsec:res_nb_test}, we have five individual detections of \SiIIf\ $\lambda$1533 halos. Three of these sources are located in \udft\ with the integration time of 31 hours, while the remaining two are in \mxdf\ with the integration times ranging from 32 hours to 141 hours (see Fig. \ref{fig:MUV_zsys} and Table \ref{table:sample}). The SB limits of the NBs used in this work depend not only on the integration time but also on the observed wavelength (i.e., redshifts of the sample). The effect of cosmic dimming is also significant and cannot be ignored. Therefore, we discuss potential effects of 1$\sigma$ SB limits and cosmic dimmings on the detectability. If the origin of \SiIIf\ emission is continuum pumping as suggested in this paper, bright UV magnitudes are necessary to produce bright \SiIIf\ emission (in terms of observed, apparent magnitudes). We examine distributions of 1$\sigma$ SB limits on a 1-arcsec$^2$ scale for the \SiIIf\ $\lambda$1533 NBs and apparent UV magnitudes for our sources in Fig. \ref{fig:muv_sb} (with integration times indicated by the color bar). We estimate the 1$\sigma$ SB limits for each object from a sensitivity table for \mxdf\ and \udft\ in \citet{bacon_muse_2023}, considering the narrow-band widths and integration times (see the left panel). The right panel shows the 1$\sigma$ SB limits multiplied by (1+$z$)$^4$/(1+$z_{1141}$)$^4$ to account for the effect of cosmic dimming in a relative manner, where $z_{1141}$ is the systemic redshift of our best-case object, ID$=1141$ ($z_{\rm sys}=2.34$). For instance, the actual depth of the \SiIIf\ NB of ID$=50$ is $3\times10^{-19}$ erg~ s$^{-1}$~cm$^{-2}$~arcsec$^{-2}$ (1$\sigma$, left panel). Additionally, because of the cosmic dimming effect, the SB level of ID$=50$ at $z_{\rm sys}=3.32$ becomes $3.3$ times fainter compared to the case where it was located at $z_{\rm sys}=2.34$. In relative terms, it corresponds to the 1$\sigma$ SB limit of $9\times10^{-19}$ erg~ s$^{-1}$~cm$^{-2}$~arcsec$^{-2}$ (right panel). Considering these two effects, the corresponding noise level of ID$=50$ is seven times higher than that of ID$=1141$, which is presented relatively as a correction in the right panel. The noise levels shown in the right panel are less dependent on the integration time (different fields) than those in the left panel. This exercise indicates that the depth of data indeed depends on the redshifts as well as integration times.

As shown in Fig. \ref{fig:muv_sb}, we cannot conclude any dependence of \SiIIf\ halo detections (red crosses) and nondetections on the SB limits as well as integration times. Even when focusing on bright objects with $m_{\rm UV}\lesssim25.5$, which are brighter than the faintest individually-detected halo (ID$=51$), and whose IDs are shown in the left panel, no clear trends are observed. The SB limits of two non-halos (ID=$6700$ at $z_{\rm sys}=2.99$ and ID=$7600$ at $z_{\rm sys}=2.57$ in \mxdf) are deeper than that of ID=$50$ at $z_{\rm sys}=3.32$ in \udft. We note that the \SiIIf\ $\lambda$1533 of one of the UV brightest objects, ID=$8048$, drops in the AO gap in the MXDF and is not included in this plot. A larger sample is needed to rigorously test the detectability depending on the depth of the dataset. In the next section, we discuss differences in sample properties. 

\FloatBarrier

\subsection{Potential differences in properties between \SiIIf-halo sources and non-halo sources}\label{ap:properties}
We check potential differences in SFR and $M_\star$ between \SiIIf-halo sources (red crosses) and non-halo sources (black circles) shown in Fig. \ref{fig:sfr_ms}, though their uncertainties are large. We find that halos tend to have high SFR and $M_\star$, which is expected from the trend with UV magnitudes (Figs. \ref{fig:MUV_zsys} and \ref{fig:muv_sb}). However, not all the sources with high SFR, high $M_\star$, or high $M_{\rm UV}$ ($m_{\rm UV}$) show a clear \SiIIf\ halo. Among them, apparent UV magnitudes (i.e., $M_{\rm UV}$ and redshifts) would be most strongly related to the detection of the halos. It is indeed consistent with the continuum pumping as the mechanism of \SiIIf\ emission. We need a larger sample to investigate the dependence of the detectability and existence of halos on data qualities and sample properties.

\end{appendix}

\end{document}